\documentclass[fleqn,usenatbib]{mnras}

\usepackage{newtxtext,newtxmath}

\usepackage[T1]{fontenc}

\DeclareRobustCommand{\VAN}[3]{#2}
\let\VANthebibliography\thebibliography
\def\thebibliography{\DeclareRobustCommand{\VAN}[3]{##3}\VANthebibliography}

\usepackage{graphicx}	
\usepackage{amsmath}	
\usepackage{xspace}
\usepackage[dvipsnames]{xcolor}

\newcommand{\OC}{$\omega$\ Cen\xspace}  
\newcommand{\SE}{$\Gamma_{\rm SE}$\xspace}   
\newcommand{\ergs}{erg s$^{-1}$\xspace}

\title[X-ray MSPs in GCs]{A Census of X-ray Millisecond Pulsars in Globular Clusters}

\author[J. Zhao \& C. O. Heinke]{
JiaQi Zhao,$^{1}$\thanks{E-mail: jzhao11@ualberta.ca}
Craig O. Heinke$^{1}$\thanks{E-mail: heinke@ualberta.ca}
\\
$^{1}$ Physics Dept., CCIS 4-183, University of Alberta, Edmonton, AB, T6G 2E1, Canada\\
}

\date{Accepted XXX. Received YYY; in original form ZZZ}

\pubyear{2015}

\begin{document}
\label{firstpage}
\pagerange{\pageref{firstpage}--\pageref{lastpage}}
\maketitle

\begin{abstract}
We present a comprehensive census of X-ray millisecond pulsars (MSPs) in 29 
Galactic globular clusters (GCs), including 
68 MSPs with confirmed X-ray luminosities and 107 MSPs with X-ray upper limits. 
We compile previous X-ray studies of GC MSPs, and add new analyses of 
six MSPs (PSRs J1326$-$4728A, J1326$-$4728B, J1518$+$0204C, J1717$+$4308A, J1737$-$0314A, and J1807$-$2459A) discovered in five GCs. Their X-ray spectra are well described by a single blackbody model, a single power-law model, or a combination of them, with X-ray luminosities ranging from 1.9$\times$10$^{30}$ \ergs to 8.3$\times$10$^{31}$ \ergs. 
We find that most detected X-ray MSPs have luminosities 
between 
$\sim 10^{30}$ erg s$^{-1}$ to $3 \times 10^{31}$ erg s$^{-1}$. 
Redback pulsars are a relatively bright MSP population with X-ray luminosities of  $\sim2\times10^{31}$--$3\times10^{32}$ \ergs. Black widows show a bi-modal distribution in X-ray luminosities, 
with 
eclipsing black widows 
between $\sim 7\times10^{30}$ and $2\times10^{31}$ \ergs,
while the two confirmed non-eclipsing black widows are much fainter, with $L_X$ of $1.5-3\times10^{30}$ \ergs, suggesting an intrinsic difference in the populations.
We estimate the total number of MSPs in 
36 GCs 
by considering the correlation between the number of MSPs and stellar encounter rate in GCs, and suggest 
that between 600--1500 MSPs exist in these 36 GCs. 
Finally, we estimate the number of X-ray detectable MSPs in the Galactic bulge, finding that 
1--86 MSPs with $L_X > 10^{33}$ erg s$^{-1}$, and 20--900 MSPs with $L_X > 10^{32}$ \ergs, should be detectable there.
\end{abstract}

\begin{keywords}
stars: neutron -- pulsars: general -- globular clusters: general -- globular clusters: individual: NGC 5139 -- X-rays: stars
\end{keywords}



\section{Introduction}

Galactic globular clusters (GCs) are ideal birthplaces for low-mass X-ray binaries (LMXBs), since the high stellar densities in GC cores provide significant chances for stellar interactions, such as tidal capture and exchange interactions \citep[see e.g.,][]{Fabian1975,Hills1976}. 
LMXBs are 
the progenitors of millisecond pulsars (MSPs), where the neutron star (NS) is spun up by accreting mass and angular momentum from its companion to a period of a few milliseconds \citep{Alpar1982,Bhattacharya1991,Papitto13}. Therefore, it is not surprising that GCs show an overabundance of MSPs compared to the Galactic field, and many MSPs exist in binary systems. 

Rotation-powered MSPs are rapidly and stably spinning pulsars, with spin periods $P \lesssim 25$ ms and spin-down rates $\Dot{P} \sim 10^{-20}$, with lifetimes of Gyrs. Apart from isolated MSPs, MSPs in binaries can be further categorized 
according to the degeneracy of the companion star. Those MSPs coupled with non-degenerate companion stars are usually referred to as `spider' MSPs, and based on the companion masses, they are further grouped as redbacks ($M_c \sim 0.2$ M$_{\sun}$, hereafter RBs) and black widows ($M_c \sim 0.02$ M$_{\sun}$, hereafter BWs), respectively. In addition, it is common that 
Eclipses of the radio emission commonly occur in observations of spider pulsars, which can be explained as the radio emission from the MSP being absorbed and/or scattered by the plasma produced at the collision between the relativistic pulsar wind and material from the companion \citep{Fruchter92,Thompson94,Stappers01,Polzin18,Zhao2020a}. On the other hand, the companion stars of MSPs may also be compact objects, and particularly white dwarfs (WDs) are the most common companions among these MSP binaries \citep[e.g.][]{Lorimer08}. 

MSPs are generally faint X-ray sources, with typical luminosities of $L_X \lesssim 10^{31}$ erg s$^{-1}$. The X-rays from MSPs commonly are observed with blackbody-like spectra \citep[e.g.][]{Zavlin02,Bogdanov2006}, indicating a thermal emission origin  likely produced from the hotspots at the NS magnetic poles, heated by the returning particles accelerated in the pulsar magnetosphere \citep[][]{Harding2002}. 
Broad, sinusoidal X-ray pulsations can be observed from nearby MSPs showing thermal emission, providing evidence for the hotspots/polar cap hypothesis \citep[see, e.g.][]{Guillot2019}. A few MSPs are relatively X-ray-bright with $L_X \gtrsim 10^{32}$ erg s$^{-1}$, up to $\sim 10^{33}$ erg s$^{-1}$, and their X-ray emission usually shows non-thermal properties (e.g. power-law spectra). For instance, the most luminous X-ray MSP in GCs found to date is PSR B1821$-$24 in M28, with $L_X = 1.4 \times 10^{33}$ erg s$^{-1}$ \citep[0.3--8 keV,][]{Saito97,Becker03,Bogdanov2011}. Moreover, the X-ray pulsations from this MSP were clearly observed in two narrow pulses \citep{Saito97,Rutledge2004}, implying  highly beamed non-thermal emission originating  from the pulsar magnetosphere. The other type of non-thermal X-ray emission from MSPs, i.e. non-pulsed non-thermal emission, is typically detected from spider pulsars, and believed to be produced by relativistic intra-binary shocks as a result of collisions between the pulsar wind and a matter outflow from the companion \citep[e.g.][]{Arons1993,Wadiasingh18,Kandel19}. 
Alternatively, a pulsar wind nebula or bow shock could also produce non-thermal X-ray emission \citep[e.g.][]{Stappers2003,Romani2017}, although pulsar wind nebulae are unlikely to be detectable (i.e. $L_X<10^{29}$ erg/s) for the spindown powers typical ($\dot{E}\sim10^{33}$ erg/s) of GC MSPs (\citealt{Kargaltsev10} show the relation between pulsar wind nebula $L_X$ and $\dot{E}$).  Bow shocks are unlikely to produce significant X-ray emission for MSPs in GCs, due to the low gas content of GCs \citep{Freire01b} and the low space velocities of MSPs in GCs.

X-ray studies of GC MSPs have been presented for a few GCs, especially  pulsar-rich clusters such as 47 Tuc \citep{Bogdanov2006}, NGC 6397 \citep{Bogdanov2010}, M28 \citep{Bogdanov2011}, NGC 6752 \citep{Forestell2014}, and recently on Terzan 5 \citep{Bogdanov2021}, M62 \citep{Oh2020} and M13 \citep{Zhao2021}, as well as for several individual MSPs in globular clusters \citep[e.g.][]{Bassa2004,Amato2019}. These studies provide opportunities to statistically investigate the X-ray properties of GC MSPs. However, different groups may apply different energy bands for spectral fitting and analysis, making it difficult to study those GC MSPs together directly. \citet{Possenti2002} used a sample of 39 pulsars, including both MSPs and normal pulsars in GCs and the Galactic field, to re-examine the correlation between X-ray and spin-down luminosities, where they converted all the fluxes to 2--10 keV. Recently, \citet{Lee2018} focused on X-ray MSPs in the Galactic field and conducted a survey of their X-ray properties. They simply applied a pure power-law model for all of the sampled MSPs and normalized the energy band to 2--10 keV as well. However, the derived X-ray luminosities of MSPs in these works may have large uncertainties, mainly due to the difficulties of measuring distances to the field MSPs \citep[e.g.][]{Igoshev16}. By contrast, the distance to a GC can be measured much more accurately than to a field MSP, and hence the uncertainty of distance to MSPs in a GC may be largely reduced \citep[now $<$5\%, e.g.][]{Baumgardt21}. 

In this paper, we present an X-ray survey for all the radio-detected MSPs in Galactic GCs which have {\it Chandra X-ray observatory} data, and which lack the confusion of bright LMXBs. We also report the X-ray spectral analyses of two newly found MSPs in the cluster Omega Centauri, using archival  {\it Chandra} observations, and statistically study the X-ray properties of GC MSPs, with particular attention to the implications for X-ray studies of the population of MSPs in the Galactic Center. This paper is organized as follows. In Section 2, we described the criteria of data collection, reduction and normalization. In Section 3, we present the results of X-ray spectral fitting for 
six new MSPs in five GCs, and catalog X-ray sources in three GCs.  We then present our X-ray census of GC MSPs, and further analyses of their X-ray properties. We discuss our results and the implications in Section 4, and we draw conclusions in Section 5.

\section{Data Collection and Reduction}
\label{sec:data}

\begin{table*}
    \centering
    \caption{Parameters for the GCs in this work.}
    \begin{tabular}{lccccccc}
    \hline
        GC Name & Distance$^a$ & {$N_{\rm H}$}$^b$ & {N$_{\rm MSP}$}$^c$ & {N$_{\rm MSP, T}$}$^d$ & Encounter Rate$^e$ & {Lim. $L_X$}$^f$ & Reference of \\
         & (kpc) & (cm$^{-2}$) & & & $\Gamma_{\rm SE}$ & (erg s$^{-1}$) & Lim. $L_X$ \\
    \hline
47 Tuc (NGC 104)	&	4.5	& $	3.48 \times 10^{20	}$ &	27 & 23	&	1000	$_{-	134	}^{+	154	}$	&	$	3 \times 10^{29	}$ & 1	\\
NGC 1851	&	12.1	& $	1.74 \times 10^{20	}$ &	13	& 1 &	1530	$_{-	186	}^{+	198	}$	&	$*$	&		\\
M53 (NGC 5024)	&	17.9	& $	1.74 \times 10^{20	}$ &	3	& 0 &	35.4	$_{-	9.6	}^{+	12.4	}$	&	$	3 \times 10^{31	}$ & 2	\\
\OC (NGC 5139)	&	5.2	& $	1.05 \times 10^{21	}$ &	5	& 2 &	90.4	$_{-	20.4	}^{+	26.3	}$	&	$	1 \times 10^{30	}$ & 3	\\
M3 (NGC 5272)	&	10.2	& $	8.71 \times 10^{19	}$ &	6	& 2 &	194.0	$_{-	18.0	}^{+	33.1	}$	&	$	2 \times 10^{31	}$ & 4	\\
M5 (NGC 5904)       	&	7.5	& $	2.61 \times 10^{20	}$ &	7	& 3 &	164.0	$_{-	0.4	}^{+	38.6	}$	&	$	5 \times 10^{30	}$ & 5	\\
NGC 5986	&	10.4	& $	2.44 \times 10^{21	}$ &	1	& 0 &	61.9	$_{-	10.4	}^{+	15.9	}$	&	$**$	&		\\
M4 (NGC 6121)	&	2.2	& $	3.05 \times 10^{21	}$ &	1	& 1 &	26.90	$_{-	9.56	}^{+	11.60	}$	&	$	3 \times 10^{29	}$ & 5	\\
M13 (NGC 6205)	&	7.4	& $	1.74 \times 10^{20	}$ &	6	& 6 &	68.9	$_{-	14.6	}^{+	18.1	}$	&	$	5 \times 10^{30	}$ & 5	\\
M12 (NGC 6218)	&	4.8	& $	1.66 \times 10^{21	}$ &	1	& 0 &	13.00	$_{-	4.03	}^{+	5.44	}$	&	$	6 \times 10^{30	}$ & 5	\\
M10 (NGC 6254)	&	4.4	& $	2.44 \times 10^{21	}$ &	2	& 0 &	31.40	$_{-	4.08	}^{+	4.34	}$	&	$	5 \times 10^{30	}$ & 5	\\
M62 (NGC 6266)	&	6.8	& $	4.09 \times 10^{21	}$ &	7	& 6 &	1670	$_{-	569	}^{+	709	}$	&	$	3 \times 10^{30	}$ & 5	\\
M92 (NGC 6341)	&	8.3	& $	1.74 \times 10^{20	}$ &	1	& 1 &	270.0	$_{-	29.0	}^{+	30.1	}$	&	$	6 \times 10^{30	}$ & 5	\\
NGC 6342	&	8.5	& $	4.01 \times 10^{21	}$ &	1	& 0 &	44.8	$_{-	12.5	}^{+	14.4	}$	&	$	5 \times 10^{31	}$ & 2	\\
Terzan 1	&	6.7	& $	1.73 \times 10^{22	}$ &	6	& 0 &	0.292	$_{-	0.170	}^{+	0.274	}$	&	$	2 \times 10^{31	}$ & 5	\\
M14 (NGC 6402)	&	9.3	& $	5.23 \times 10^{21	}$ &	5	& 1 &	124.0	$_{-	30.2	}^{+	31.8	}$	&	$	6 \times 10^{31	}$ & 5	\\
NGC 6397	&	2.3	& $	1.57 \times 10^{21	}$ &	2	& 2 &	84.1	$_{-	18.3	}^{+	18.3	}$	&	$	1 \times 10^{29	}$ & 5	\\
Terzan 5	&	6.9	& $	1.99 \times 10^{22	}$ &	38	& 37 &	6800	$_{-	3020	}^{+	1040	}$	&	$	1 \times 10^{30	}$ & 5	\\
NGC 6440	&	8.5	& $	9.32 \times 10^{21	}$ &	7	& 5 &	1400	$_{-	477	}^{+	628	}$	&	$	4 \times 10^{31	}$ & 6	\\
NGC 6441	&	11.6	& $	4.09 \times 10^{21	}$ &	5	& 3 & 	2300	$_{-	635	}^{+	974	}$	&	$*$	&		\\
NGC 6517	&	10.6	& $	9.41 \times 10^{21	}$ &	8	& 6 &	338.0	$_{-	97.5	}^{+	152.0	}$	&	$	3 \times 10^{31	}$ & 2	\\
NGC 6522	&	7.7	& $	4.18 \times 10^{21	}$ &	4	& 1 &	363.0	$_{-	98.5	}^{+	113.0	}$	&	$	5 \times 10^{30	}$ & 5	\\
NGC 6539	&	7.8	& $	8.89 \times 10^{21	}$ &	1	& 1 &	42.1	$_{-	15.3	}^{+	28.6	}$	&	$	5 \times 10^{31	}$ & 5	\\
NGC 6544	&	3.0	& $	6.62 \times 10^{21	}$ &	2	& 2 &	111.0	$_{-	36.5	}^{+	67.8	}$	&	$	6 \times 10^{30	}$ & 5	\\
NGC 6624	&	7.9	& $	2.44 \times 10^{21	}$ &	9	& 3 &	1150	$_{-	178	}^{+	113	}$	&	$*$	&		\\
M28 (NGC 6626)	&	5.5	& $	3.48 \times 10^{21	}$ &	13	& 8 &	648.0	$_{-	91.1	}^{+	83.8	}$	&	$	8 \times 10^{29	}$ & 5	\\
NGC 6652	&	10.0	& $	7.84 \times 10^{20	}$ &	2	& 0 &	700	$_{-	189	}^{+	292	}$	&	$	2 \times 10^{31	}$ & 7	\\
M22 (NGC 6656)	&	3.2	& $	2.96 \times 10^{21	}$ &	2	& 2 &	77.5	$_{-	25.9	}^{+	31.9	}$	&	$	8 \times 10^{29	}$ & 5	\\
NGC 6712	&	6.9	& $	3.92 \times 10^{21	}$ &	1	& 1 &	30.80	$_{-	6.64	}^{+	5.63	}$	&	$*$	&		\\
NGC 6749	&	7.9	& $	1.31 \times 10^{22	}$ &	2	& 1 &	51.5	$_{-	20.9	}^{+	40.7	}$	&	$**$	&		\\
NGC 6752	&	4.0	& $	3.48 \times 10^{20	}$ &	9	& 6 &	401	$_{-	126	}^{+	182	}$	&	$	3 \times 10^{29	}$ & 8	\\
NGC 6760	&	7.4	& $	6.71 \times 10^{21	}$ &	2	& 2 &	56.9	$_{-	19.4	}^{+	26.6	}$	&	$	1 \times 10^{31	}$ & 5	\\
M71 (NGC 6838)	&	4.0	& $	2.18 \times 10^{21	}$ &	2	& 1 &	1.470	$_{-	0.138	}^{+	0.146	}$	&	$	2 \times 10^{30	}$ & 9	\\
M15 (NGC 7078)	&	10.4	& $	8.71 \times 10^{20	}$ &	5	& 4 &	4510	$_{-	986	}^{+	1360	}$	&	$*$	&		\\
M2 (NGC 7089)	&	11.5	& $	5.23 \times 10^{20	}$ &	5	& 0 &	518.0	$_{-	71.4	}^{+	77.6	}$	&	$	6 \times 10^{31	}$ & 5	\\
M30 (NGC 7099)	&	8.1	& $	2.61 \times 10^{20	}$ &	2	& 1 &	324.0	$_{-	81.2	}^{+	124.0	}$	&	$	2 \times 10^{30	}$ & 5	\\
    \hline
    \multicolumn{8}{l}{{\it Notes}: $^a$ Distance to GCs collected from \citet[2010 edition]{Harris1996}.  } \\
    \multicolumn{8}{p{14cm}}{$^b$ Hydrogen column number density towards GCs calculated based on correlation between $N_{\rm H}$ and optical extinction, $A_V$ \citep{Bahramian2015}.} \\
    \multicolumn{8}{l}{$^c$ Number of discovered MSPs. } \\
    \multicolumn{8}{l}{$^d$ Number of discovered MSPs with precise timing positions. } \\
    \multicolumn{8}{l}{$^e$ Stellar encounter rate \SE estimated by \citet{Bahramian2013}, with 1-$\sigma$ errors. } \\
    \multicolumn{8}{p{14cm}}{$^f$ The limiting unabsorbed X-ray luminosity estimated in the band 0.3--8 keV. Reference: (1) \citet{Cheng2019}; (2) this work (see section~\ref{subsec:limLx}); (3) \citet{Henleywillis2018}; (4) \citet{Zhao2019}; (5) \citet{Bahramian2020}; (6) \citet{Pooley2002};  (7) \citet{Stacey2012}; (8) \citet{Forestell2014}; (9) \citet{Elsner2008}.} \\
    \multicolumn{8}{l}{$*$ Severely contaminated by bright X-ray sources \citep[see][]{Verbunt2006}.} \\
    \multicolumn{8}{l}{$**$ No CXO observations} \\
    \end{tabular}
    \label{tab:parameters}
\end{table*}

Based on the catalogue of pulsars in globular clusters (230 pulsars in 36 GCs to date)\footnote{\url{http://www.naic.edu/~pfreire/GCpsr.html}}, we first produced a list of MSPs in GCs by defining their spinning periods of $P \lesssim 25$ ms. The boundary of rotational periods between normal pulsars and MSPs is not solid and may vary depending on the research of interests. Here we chose $\sim 25$ ms as the upper limit. 
This initial filtering gave 210 MSPs in total, and each GC in the catalogue harbours at least one MSP. Table~\ref{tab:parameters} lists all the 36 GCs studied in this work, as well as other parameters. 

To obtain the X-ray luminosities of these MSPs, we 
looked into {\it Chandra X-ray Observatory} (CXO) observations. 
Except NGC 5986 and NGC 6749, which have no CXO observation yet, all GCs have been observed at least once 
with 
a total exposure time  $>$10 ks. However, five GCs (NGC 1851, NGC 6441, NGC 6624, NGC 6712, and M15) are 
not feasible for the X-ray analysis of 
MSPs or other faint X-ray sources, given that one or more bright ($L_X>10^{36}$ erg/s) XRBs are 
present, producing a high X-ray background throughout the cluster core 
\citep[see][for a review of GC X-ray sources]{Verbunt2006}. Hence there remain 29 GCs 
where we can 
determine or constrain the X-ray luminosities of known MSPs. Several GCs and the MSPs therein have been observed and studied in X-rays thoroughly, such as 47 Tuc, M28, etc. (see Introduction), 
while other GCs have had deep surveys of X-ray sources before MSPs were detected therein  \citep[e.g.][]{Henleywillis2018}. Also, \citet{Bahramian2020} provided a comprehensive catalogue of faint X-ray sources in 38 GCs, which may contain information about MSPs. 
In addition, to test the robustness of X-ray luminosities derived by \citet{Bahramian2020}, we collected a dozen MSPs with well determined X-ray luminosities in the literature, and compared the literature values with the corresponding X-ray luminosities in the Bahramian catalogue. We found the values agree within their errors, and hence the X-ray luminosities presented by \citet{Bahramian2020} appear  reliable. Three GCs (M53, NGC 6342, and NGC 6517), however, do not have published X-ray surveys yet, though archival CXO data is available on them. Therefore, we collected and extracted X-ray information (e.g. fluxes and luminosities) of GC MSPs based on previous studies of them. 
To normalize the X-ray energy range, we use unabsorbed X-ray luminosities in the 0.3--8 keV band as the normalization, and all the X-ray MSPs and sources studied in other energy bands were converted into the 0.3--8 keV band via the {\it Chandra} proposal planning tool, {\sc pimms}\footnote{\url{https://asc.harvard.edu/toolkit/pimms.jsp}}.
Generally, the errors introduced by assuming a spectrum to convert bands (e.g. 0.5-2 to 0.3-8) in \textsc{pimms} are $<$ 20\%, if $N_H$ does not exceed $10^{22}$ cm$^{-2}$. These errors, however, 
are generally small when compared with the 
statistical errors of low count sources\footnote{See \url{https://cxc.harvard.edu/ciao/why/pimms.html} for more details.}, typical of MSPs in this work. 
The choice of energy band 0.3--8 keV emphasizes the X-ray emission from MSPs, including both thermal and non-thermal X-rays.

\subsection{GC MSPs with X-ray analysis}

A few GCs contain a large number of MSPs, like 38 MSPs found in Terzan 5 and 27 MSPs found in 47 Tuc, and hence they are of great interest to study the X-ray properties of GC MSPs. Given the deep X-ray observations by CXO as well as radio timing observations of these MSPs, their X-ray spectra may be well extracted and modeled, and therefore we can simply obtain their X-ray luminosities and other properties from corresponding studies. For instance, there are 20 MSPs in 47 Tuc that have spectral analysis with well fitted models and unabsorbed luminosities \citep[see][]{Bogdanov2006, Bhattacharya2017}. Similarly, most MSPs in M13 and NGC 6752, and many in Terzan 5 and M28, also have well determined X-ray luminosities \citep[][]{Bogdanov2011,Forestell2014,Linares2014,Bogdanov2021,Zhao2021}. 

While new MSPs are continuously being discovered in GCs \citep[e.g.][]{Pan2021b,Ridolfi2021}, 
most do not yet have precise timing positions.
Alternatively, we can constrain the X-ray luminosities for 
MSPs 
without precise timing positions by setting  upper limits using the known cluster X-ray sources. Given the 
detailed multiwavelength analyses of several 
GCs \citep[e.g.][]{Pooley2002,Edmonds03a,Heinke2005}, most X-ray sources with $L_X \gtrsim 3 \times 10^{32}\ {\rm erg\ s^{-1}}$ have been identified as CVs, LMXBs, etc, and only faint X-ray sources ($L_X \lesssim 10^{32}\ {\rm erg\ s^{-1}}$)
often remain unidentified. 
Since MSPs are typically faint X-ray emitters, we can then define the X-ray luminosity upper limit of those MSPs without X-ray identifications in one GC as the luminosity of the brightest unidentified X-ray source in that GC. 
Hence we simply use one upper limit of X-ray luminosity for each cluster for all the 
MSPs without known positions in that cluster.
While for those MSPs with published timing positions but not studied yet in X-rays, we briefly look into their X-ray luminosities and place upper limits for them (see below).
It is a conservative definition but fair enough to give us a sense of the X-ray brightness of those MSPs. 

We note that X-ray studies of GC sources generally assume a single power-law spectrum with a fixed photon index $\Gamma$ for all the detected faint sources \citep[e.g.][]{Cackett2006,Henleywillis2018}, and consequently the fitted parameters might not reflect their intrinsic X-ray properties. 




\subsection{GC X-ray analysis in this work}

We briefly analyze CXO observations of three GCs (M53, NGC 6342, and NGC 6517) that have not been studied yet to constrain the X-ray luminosities of the sources therein. Also, we analyze X-ray spectra of 
six MSPs with clear X-ray counterparts at their precise timing positions in five GCs (\OC, M5, M92, M14, and NGC 6544).
In addition, we extract X-ray flux limits for those MSPs with timing positions but 
without an X-ray counterpart
to calculate their upper limits of X-ray luminosities (
25 MSPs in 10 GCs; see Table~\ref{tab:appendix_table}). 
We note that MSPs with new timing positions in the cluster M62 
will require
detailed X-ray studies,
as the potential counterparts are located in a crowded core where careful astrometric considerations will be necessary (see e.g. \citealt{Bogdanov2021} for a similar analysis of Terzan 5).
Therefore, we only give upper limits of their X-ray luminosities in this paper, and will study them carefully in future work.
The data reduction and analysis were performed using {\sc ciao}\footnote{Chandra Interactive Analysis of Observations, available at \url{https://cxc.cfa.harvard.edu/ciao/}.} \citep[version 4.13, {\sc caldb} 4.9.4,][]{Fruscione2006}. All the {\it Chandra} observation data were first reprocessed using the {\tt chandra\_repro} script to create new level 2 event files that apply the latest calibration updates and bad pixel files. Plus, we filtered the data to the energy band 0.3--8 keV to keep consistency.  No background flares were detected in the CXO observations, 
except for M92 and NGC 6544 (see below). 

Furthermore, we checked the astrometry between the X-ray images and the radio timing positions for the clusters where we perform X-ray analysis.
We first ran the {\tt wavdetect} script, a Mexican-Hat Wavelet source detection algorithm\footnote{\url{https://cxc.cfa.harvard.edu/ciao/threads/wavdetect/}}, to get the X-ray positions of bright sources on the S3 chip (the aimpoint chip for most {\it Chandra} observations), or the I0-I3 chips for the \OC observations. Then we checked {\it Gaia} Data Release 3 \citep[DR3;][]{GaiaCollaboration2016,GaiaCollaboration2021} to see if there are one or more bright {\it Gaia} stars (with G-band magnitude $<$ 16 mag to reduce the number of chance coincidences) within 1\arcsec\ of the {\it Chandra} sources, outside the cluster half-light radius. We applied the mean value of the offsets between {\it Chandra} and {\it Gaia} detections in each cluster to MSP positions to correct absolute astrometry, and used the corrected positions for the following analysis. We found the offsets are (+0.110\arcsec, $-$0.125\arcsec), (+0.194\arcsec, $-$0.017\arcsec), ($-$0.011\arcsec, $-$0.163\arcsec), and (+0.012\arcsec, +0.052\arcsec) for the clusters \OC, M92, M14, and NGC 6544, respectively, in R.A. and Dec. 

For the cluster M5, however, we could not find any eligible {\it Chandra-Gaia} matches to refine the astrometric alignment. However, based on the radio timing position of PSR J1518+0204C (or M5C) in M5, we found its X-ray counterpart on the {\it Chandra} image via {\tt wavdetect}, with a radial offset of $\sim$0.4\arcsec\ to the radio position. Since the overall 90\% uncertainty circle of {\it Chandra} X-ray absolute position has a radius of 0.8\arcsec,\footnote{see \url{https://cxc.harvard.edu/cal/ASPECT/celmon/}} and the probability of a spurious association is low (about 1\% chance of the pulsar lying within 1\arcsec of one of the 8 X-ray sources in the core), we argue that the {\it Chandra} absolute astrometric accuracy for M5 is satisfactory for the following analysis.

\subsubsection{M53, NGC 6342, and NGC 6517}
\label{subsubsec:src_det}

\begin{figure*}
    \centering
    \includegraphics[width=\textwidth]{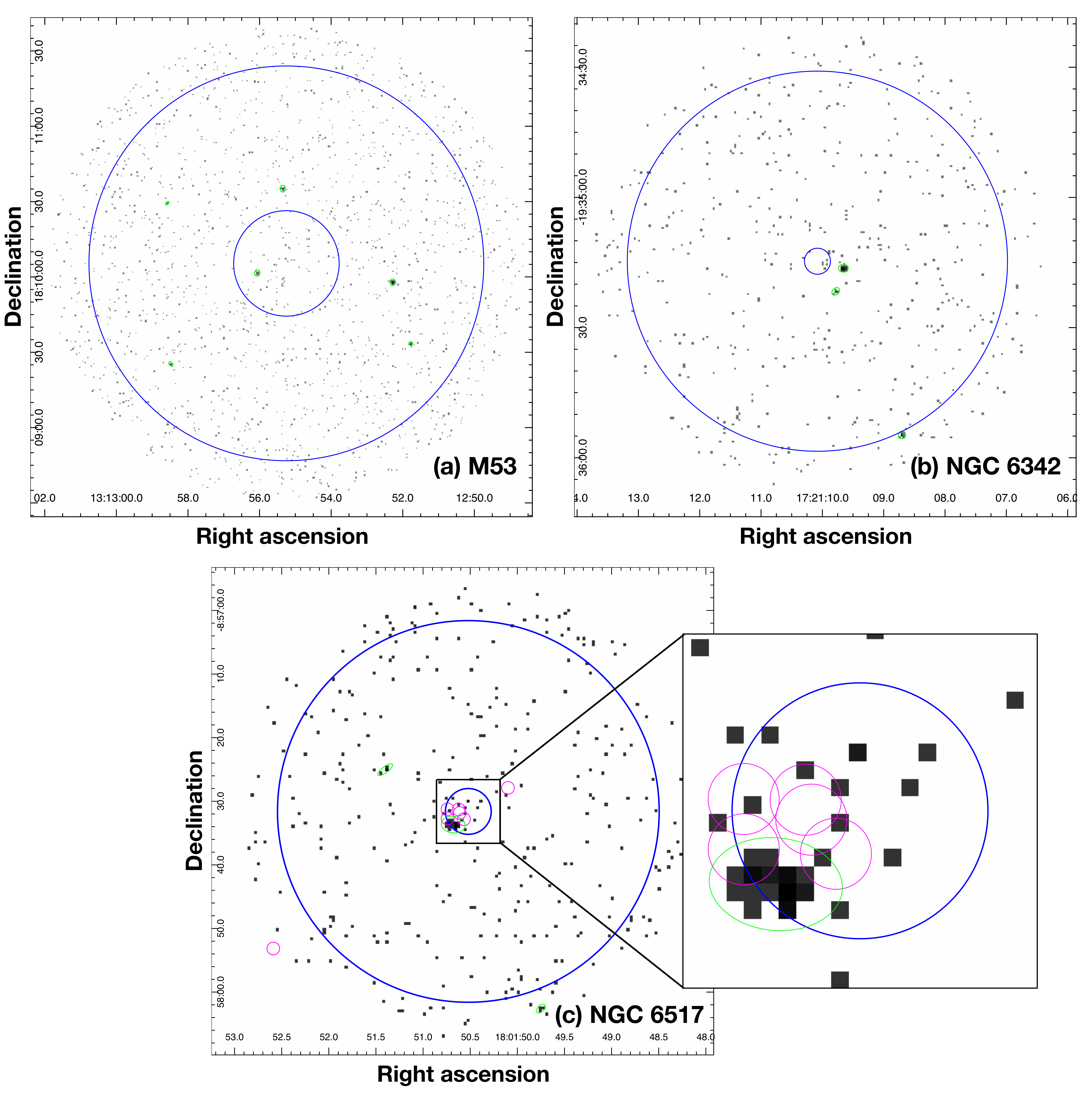}
    \caption{{\it Chandra} X-ray images in the band 0.3--8 keV of M53, or NGC 5024 (a); NGC 6342 (b); and NGC 6517 (c), respectively. The smaller blue circle in each panel shows the core region of the corresponding GC with a radius of 0.35\arcmin, 0.05\arcmin, and 0.06\arcmin, respectively, while the larger blue circle shows the half-light radius of 1.31\arcmin, 0.73\arcmin, and 0.50\arcmin, respectively \citep[][2010 edition]{Harris1996}. Green contours indicate X-ray sources detected by {\tt wavdetect} (see text for details). For each cluster, the GC region searched was defined as 1.2 times the half-light radius. North is up, and east is to the left. For NGC 6517, we also marked out the MSPs therein with known timing positions using magenta circles with 1\arcsec\ radii, while the zoomed-in figure shows the central 10\arcsec\ $\times$ 10\arcsec\ region. We note that PSR J1801$-$0857D in NGC 6517 has timing positions though, it is distant from  the centre with an offset of 1.2\arcmin and hence not included in this figure.}
    \label{fig:3_X-ray_images}
\end{figure*}

Each of M53, NGC 6342, and NGC 6517 has been observed once by CXO in the VFAINT mode, with exposure times ranging from 15 ks to 25 ks (see Table~\ref{tab:obs_GCs}).
We formally detected X-ray sources in each GC by using {\tt wavdetect}. We set the wavelet scales of 1.0, 1.4, 2.0, and 4.0, and the significance threshold of $10^{-6}$ (false positives per pixel). The detected X-ray sources are shown in Figure~\ref{fig:3_X-ray_images} (green contours). To generate the X-ray luminosities of those sources, we first 
extracted their spectra from the detection regions provided by wavdetect (shown in Fig. 1) running {\tt dmextract} in {\sc ds9}. 
(We also tried to extract spectra using 1\arcsec\ radii circular regions and found that the results changed little, e.g. the spectral results remained within the errors. Hence we used the regions generated by \texttt{wavdetect} for convenience.)
With additional known observation information (e.g. {\it Chandra} cycles, detectors), we converted count rates into unabsorbed X-ray fluxes and luminosities via {\sc pimms}, assuming a power-law spectrum with a photon index of 1.7 for those sources, and distances from \citet[2010 revision]{Harris1996}. 
The choice of a power-law model with a photon index of 1.7 is a compromise that is typical of background AGN \citep{Giacconi01}, and of globular cluster X-ray sources at $L_X\sim10^{31}$ erg s$^{-1}$ \citep[e.g.][their Fig. 4]{Zhao2020b}, typical for these observations (see Table 5).

\begin{table}
	\centering
	\caption{\textit{Chandra} observations of M53, NGC 6342, and NGC 6517}
	\label{tab:obs_GCs}
	\begin{tabular}{lcccr} 
		\hline
		 & & Date of & Observation & Exposure\\
		GC Name & Instrument & Observation & ID & Time (ks) \\
		\hline
		M53 & ACIS-S & 2006 Nov 13 & 6560 & 24.5 \\
        NGC 6342 & ACIS-S & 2009 Jul 10 & 9957 & 15.8 \\
        NGC 6517 & ACIS-S & 2009 Feb 04 & 9597 & 23.6 \\
		\hline
	\end{tabular}
\end{table}

\subsubsection{\OC}

\begin{figure*}
    \centering
    \includegraphics[width=\textwidth]{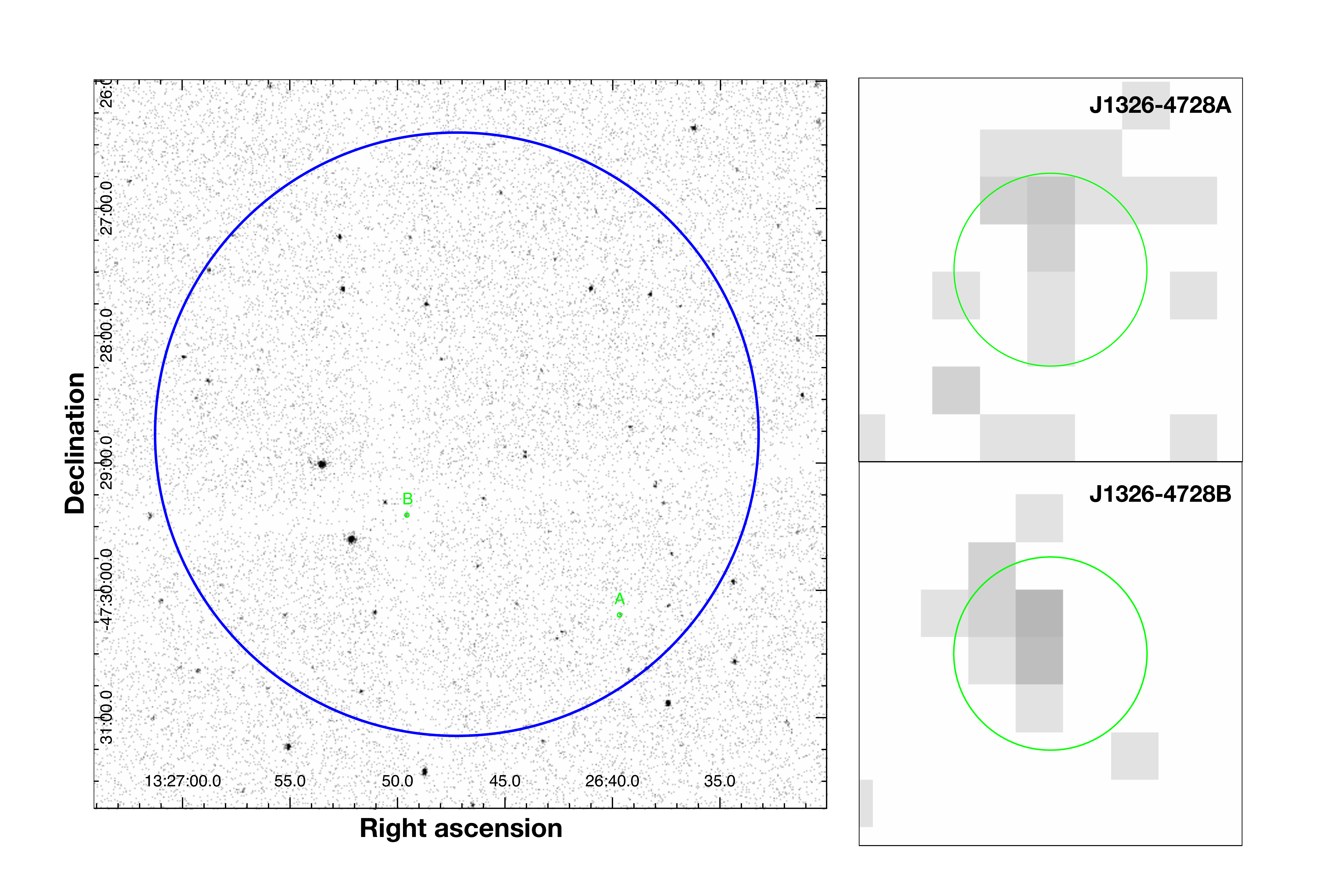}
    \caption{Merged {\it Chandra} X-ray image of \OC in the band of 0.3--8 keV. {\it Left}: The blue circle shows the core region of \OC centred at R.A. = 13:26:47.24, Dec. = $-$47:28:46.5 with a radius of 2.37\arcmin \citep[][2010 edition]{Harris1996}. Two MSPs (J1326$-$4728A and J1326$-$4728B) with timing positions are marked using green circles with 1-arcsec radii. North is up, and east is to the left. Bright X-ray sources visible in this image contain CVs \citep[see][]{Cool13,Henleywillis2018}. {\it Right}: zoomed-in X-ray images of J1326$-$4728A (upper panel) and J1326$-$4728B (lower panel), respectively, in $4\arcsec \times 4\arcsec$ boxes. X-ray emission from these two MSPs are seen clearly. }
    \label{fig:oc}
\end{figure*}

The X-ray data of \OC used in this work consist of four CXO observations, with a total exposure time of 290.1 ks (see Table~\ref{tab:obs_OC}). All of the four observations were imaged using the ACIS-I imaging array and configured in VFAINT mode. We created a co-added X-ray image of \OC in 0.3--8 keV band by merging four level 2 event files using {\tt merge\_obs} (Figure~\ref{fig:oc}).

Since \OC has had a deep and comprehensive X-ray study recently by \citet{Henleywillis2018} (see also \citealt{Cool13}), we only focus on the MSPs in \OC  discovered since  their work. There are five MSPs found to date in \OC \citep{Dai2020}, and only two of them (J1326$-$4728A and J1326$-$4728B) have precise timing positions. To analyse the X-ray spectra and obtain luminosities of these two MSPs, we first extracted X-ray emission from the circular regions with a 1-arcsec radius centred on the radio timing positions (green circles in Figure~\ref{fig:oc}) by using {\tt specextract}. 
We performed the extraction process for each observation separately, and then combined the spectra for each MSP correspondingly using {\tt combine\_spectra}. The background was taken from source-free annular areas around the MSPs.

\begin{table}
	\centering
	\caption{\textit{Chandra} observations used for MSP spectral analysis}
	\label{tab:obs_OC}
	\begin{tabular}{lcccr} 
		\hline
		& & Date of & Observation & Exposure\\
		GC Name & Instrument & Observation & ID & Time (ks) \\
		\hline
		\OC & ACIS-I & 2000 Jan 24 & 653 & 25.0 \\
         & ACIS-I & 2000 Jan 25 & 1519 & 43.6 \\
         & ACIS-I & 2012 Apr 17 & 13726 & 173.7 \\
         & ACIS-I & 2012 Apr 16 & 13727 & 48.5 \\
        M5 & ACIS-S & 2002 Sept 24 & 2676 & 44.7 \\
        M92 & ACIS-S & 2003 Oct 05 & 3778 & 29.7 \\
         & ACIS-S & 2003 Oct 19 & 5241 & 22.9 \\
        M14 & ACIS-S & 2008 May 24 & 8947 & 12.1 \\
        NGC 6544 & ACIS-S & 2005 Jul 20 & 5435 & 16.3 \\
		\hline
	\end{tabular}
\end{table}

\subsubsection{M5 (NGC 5904)}

\begin{figure}
    \centering
    \includegraphics[width=\linewidth]{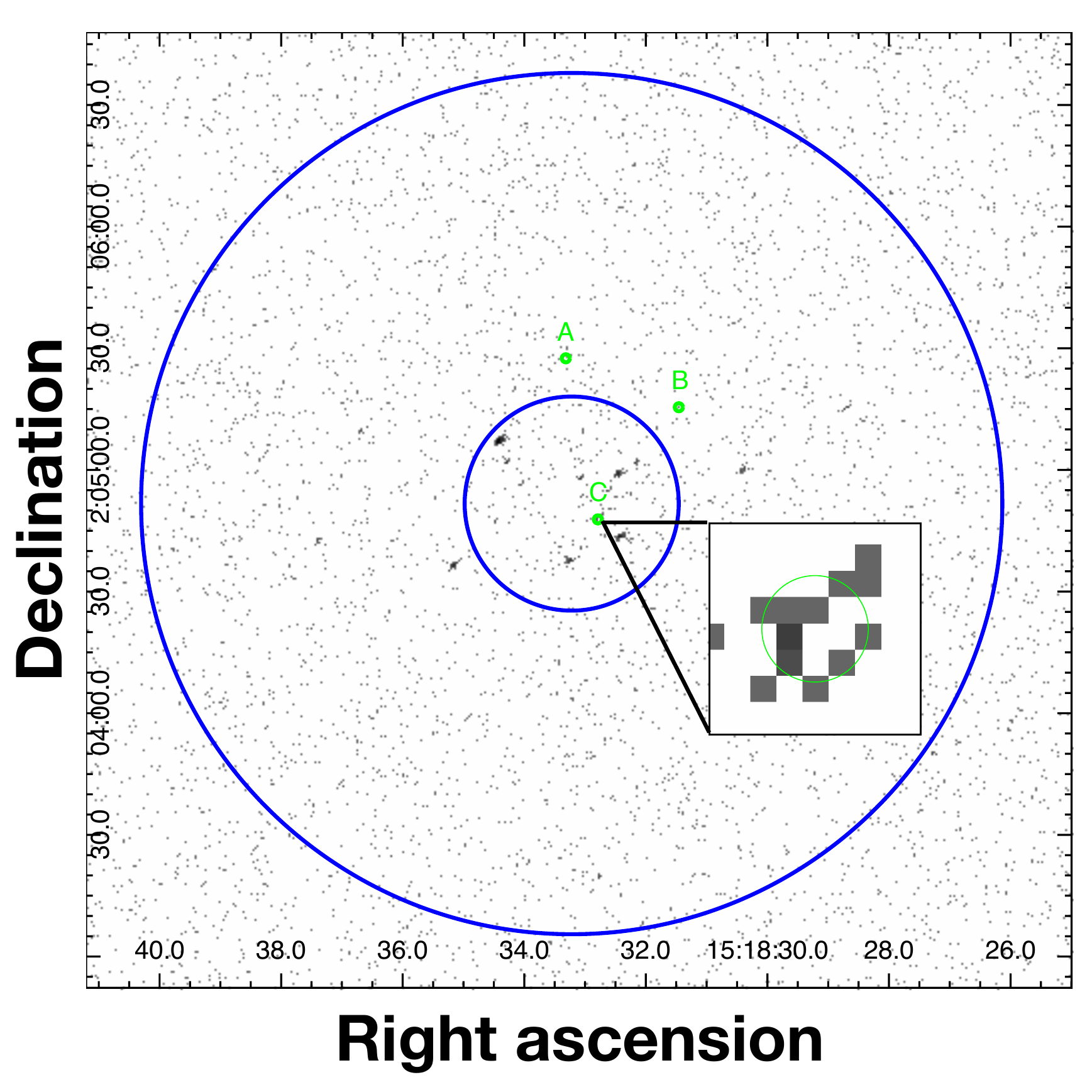}
    \caption{{\it Chandra} X-ray image of M5 in the band 0.3--8 keV. The smaller blue circle shows the 0.44\arcmin\ core radius of M5, while the larger one shows the 1.77\arcmin\ half-light radius, and both are centred at R.A. = 15:18:33.22, Dec. = +02:04:51.7 \citep[][2010 edition]{Harris1996}. North is up, and east is to the left. Three MSPs (J1518+0204A, J1518+0204B, and J1518+0204C) with timing positions are labeled with green circles with 1\arcsec\ radii, while the inset 4\arcsec\ $\times$ 4\arcsec\ box shows the zoomed-in image of J1518+0204C .}
    \label{fig:M5}
\end{figure}

M5 has been observed once by CXO with an exposure time of 44.7 ks in the FAINT data mode. Figure~\ref{fig:M5} shows the X-ray image of M5 in the band 0.3--8 keV. Seven MSPs have been found in this cluster so far \citep[][ and another, yet unpublished, by FAST, \footnote{\url{https://crafts.bao.ac.cn/pulsar/SP2/}}]{Wolszczan89,Anderson1997,Hessels07,Pan2021a}, while only three of them (J1518+0204A, J1518+0204B, and J1518+0204C) have precise timing positions (marked by green circles in Figure~\ref{fig:M5}; \citealt{Anderson1997,Pallanca2014}). Among these three MSPs, J1518+0204A 
shows a clear X-ray detection on the {\it Chandra} image (see the inset box in Figure~\ref{fig:M5}), and its counterpart is detected by {\tt wavdetect}. Hence we are able to perform X-ray spectral analysis for J1518+0204A. We note that no individual X-ray study of M5 has yet been published, while \citet{Bahramian2020} presented the X-ray source detections and source properties of this cluster in a general way. However, the X-ray counterpart to J1518+0204A is not included in their source catalogue.

\subsubsection{M92 (NGC 6341)}

\begin{figure}
    \centering
    \includegraphics[width=\linewidth]{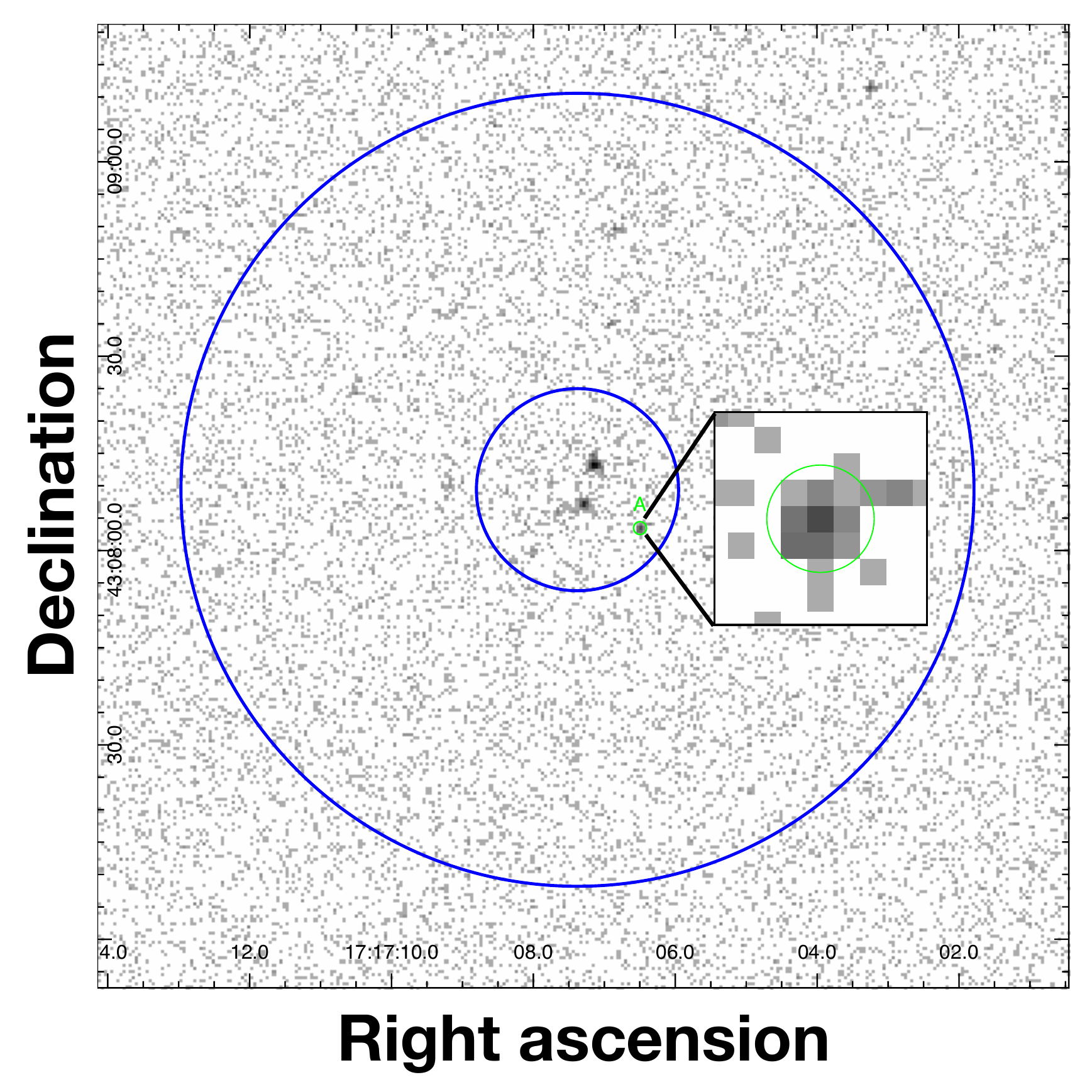}
    \caption{Merged {\it Chandra} X-ray image of M92 in the band 0.3--8 keV. The smaller blue circle shows the 0.26\arcmin\ core radius, while the larger one shows the 1.02\arcmin\ half-light radius of M92, and both are centred at R.A. = 17:17:07.39, Dec. = +43:08:09.4 \citep[][2010 edition]{Harris1996}. North is up, and east is to the left. The only discovered MSP (J1717+4308A) in M92 is marked with a 1\arcsec-radius green circle, while the inset 4\arcsec\ $\times$ 4\arcsec\ box shows the zoomed-in image of J1717+4308A. Other bright X-ray sources in the image contain CVs \citep[see][]{Lu2011}.}
    \label{fig:M92}
\end{figure}

There are two CXO observations of M92 so far (see Table~\ref{tab:obs_OC}) with a total exposure time of $\sim$52.6 ks. \citet{Lu2011} identified a few background flares during the observations, and reduced the effective exposure time to 52.5 ks. However, we argue that the short flare background has little influence on the spectral analysis of the MSP in this cluster after extracting and deducting the background emission. Hence we skipped the process of eliminating background flare times for M92, and the merged X-ray image in the band 0.3--8 keV is shown in Figure~\ref{fig:M92}.

One MSP (PSR J1717+4308A) has been discovered in M92, which is identified as an eclipsing redback \citep{Pan2020,Pan2021b}. Its X-ray counterpart is clearly seen on the {\it Chandra} X-ray image (see Figure~\ref{fig:M92}). \citet{Lu2011} comprehensively studied the X-ray sources and corresponding optical counterparts in M92, including the X-ray counterpart to J1717+4308A (source ID CX3 in their catalogue), while the X-ray spectral analysis of J1717+4308A was not contained in their work. Here, we perform an X-ray spectral analysis for J1717+4308A. 

\subsubsection{M14 (NGC 6402)}
\label{subsubsec:M14}
\begin{figure}
    \centering
    \includegraphics[width=\linewidth]{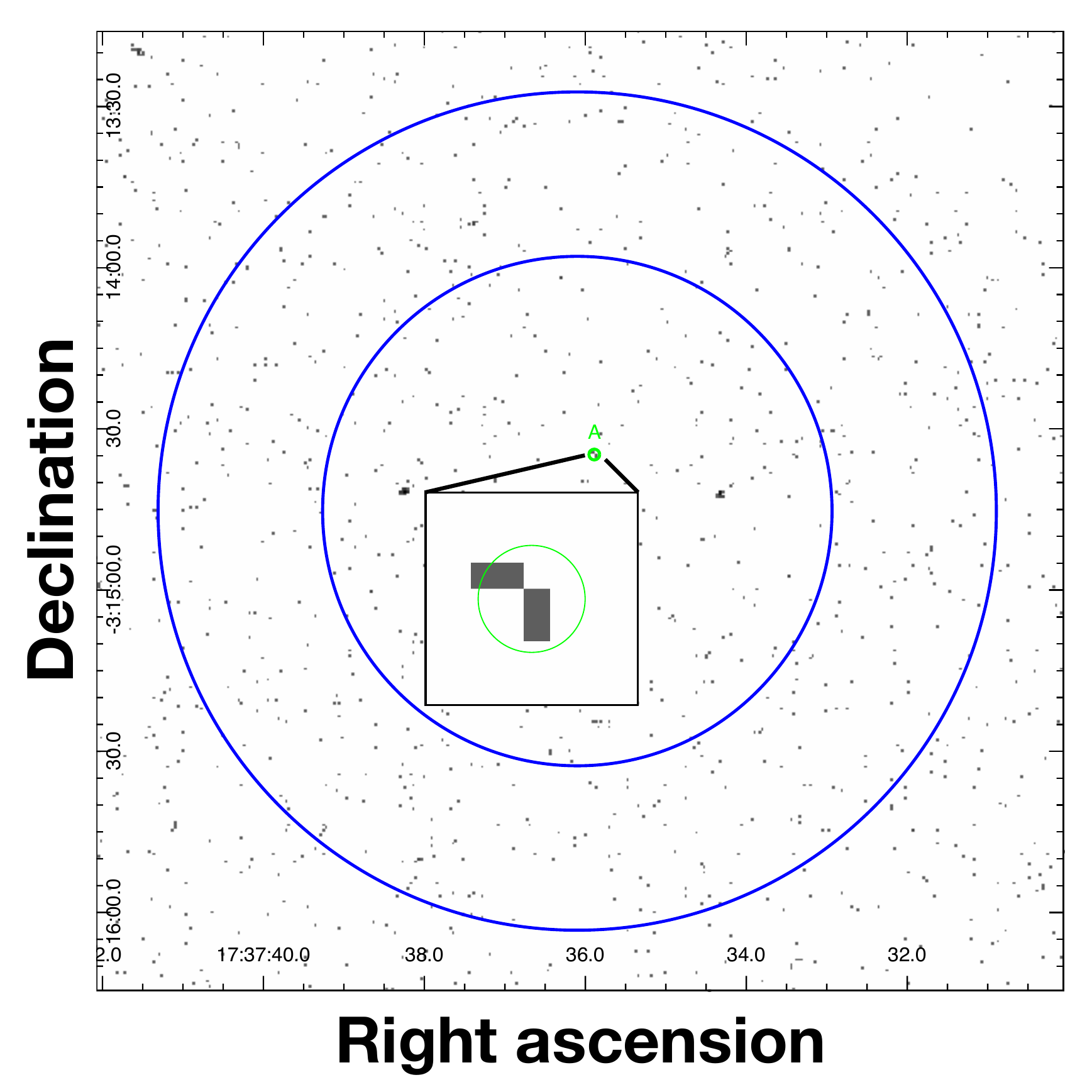}
    \caption{{\it Chandra} X-ray image in the 0.3--8 keV band of M14. The smaller and larger circles show the cluster's core radius (0.79\arcmin) and half-light radius (1.30\arcmin), respectively, and both circles are centred at R.A. = 17:37:36.10, Dec. = $-$03:14:45.3 \citep[][2010 edition]{Harris1996}. North is up, and east is to the left. PSR J1737$-$0314A in M14 is marked with a 1\arcsec-radius green circle, while the inset 4\arcsec\ $\times$ 4\arcsec\ box shows the zoomed-in image of J1737$-$0314A. }
    \label{fig:M14}
\end{figure}

M14 only has a 12.1-ks CXO observation in VFAINT data mode to date. Due to the limited exposure time and high hydrogen column density (see Table~\ref{tab:parameters}) towards this cluster, a deep X-ray study of M14 has not yet been performed. Nonetheless, \citet{Bahramian2020} presented a total of seven X-ray sources detected in this cluster. 
Five MSPs have been found recently in M14 by \citet{Pan2021b}, although only PSR J1737$-$0314A has a published precise timing solution. 
The timing solution for PSR J1737$-$0314A indicates a black widow system, but the discovery observation  was not long enough to exclude the possibility of eclipses, given the orbital period of $\sim$5.5 hours with an observation length of 2 hours \citep{Pan2021b}. Hence it is not clear if this pulsar is eclipsing. 
None of the X-ray sources published by \citet{Bahramian2020} are the counterpart to J1737$-$0314A, though there appears to be X-ray emission at its timing position. In this work, we extract and analyze the X-ray spectrum of J1737$-$0314A based on its timing position. The {\it Chandra} X-ray image and X-ray emission from J1737$-$0314A can be found in Figure~\ref{fig:M14}. 

\subsubsection{NGC 6544}

\begin{figure}
    \centering
    \includegraphics[width=\linewidth]{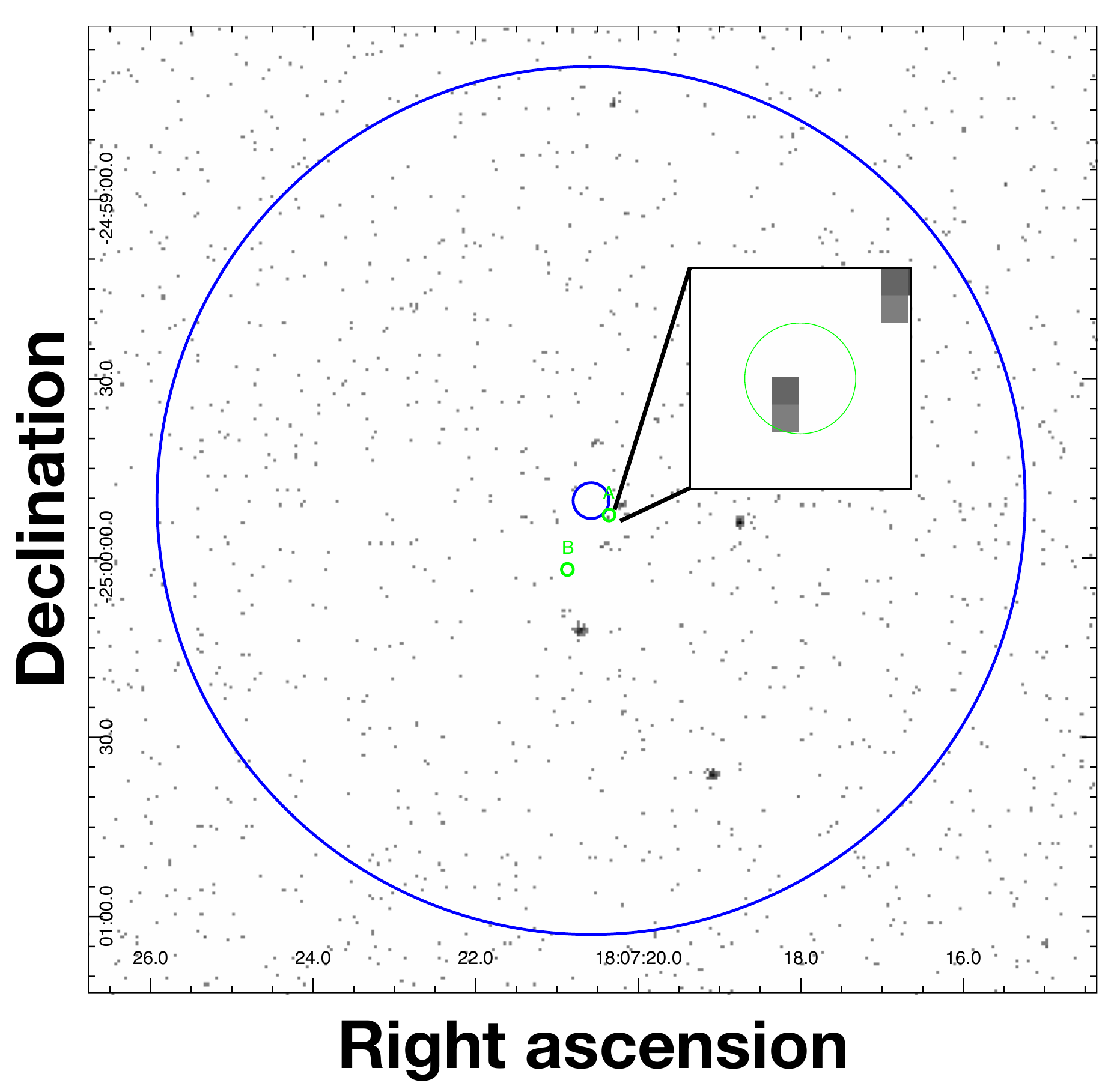}
    \caption{{\it Chandra} X-ray image in the band 0.3--8 keV of NGC 6544. The smaller and larger circles show core radius (0.05\arcmin) and half-light radius (1.21\arcmin), respectively, and both circles are centred at R.A. = 18:07:20.58, Dec. = $-$24:59:50.4 \citep[][2010 edition]{Harris1996}. North is up, and east is to the left. PSR J1807$-$2459A in NGC6544 is marked with an 1\arcsec-radius green circle, while the inset 4\arcsec\ $\times$ 4\arcsec\ box shows the zoomed-in image of J1807$-$2459A. }
    \label{fig:NGC6544}
\end{figure}

There is one CXO observation of NGC 6544, with $\sim$16.3 ks exposure time and FAINT data configuration. 
Since NGC 6544 is a cluster with substantial background flaring,  we eliminated the high background times by limiting the count rate $<$15 counts s$^{-1}$ from the entire chip S3, resulting in an effective exposure time of $\sim$12.7 ks. 
NGC 6544 is a so-called core-collapsed GC, with a core radius of only  0.05\arcsec\ \citep[][2010 edition]{Harris1996}. Two MSPs (J1807$-$2459A and J1807$-$2459B) with timing positions have been discovered in this GC \citep{DAmico01,Ransom2001,Lynch2012}, while only J1807$-$2459A shows a clear X-ray counterpart (see Figure~\ref{fig:NGC6544}). \citet{Bahramian2020} published eight X-ray sources in NGC 6544, although the X-ray counterpart to J1807$-$2459A is not included. We perform X-ray analysis for this MSP in this paper. 

\subsubsection{Other MSPs with timing positions}

There is another group of MSPs that have radio timing positions 
but have no or few X-ray counts from corresponding regions, like the MSPs in NGC 6517 (see Figure~\ref{fig:3_X-ray_images}.c).
In this work, we briefly investigate their X-ray properties to further constrain the X-ray luminosity distribution of GC MSPs. For these MSPs, we apply CXO observations of their corresponding GCs and extract their X-ray fluxes within a circle with 1-arcsec radius and centred at their timing positions. We use {\tt srcflux} in {\sc ds9} to calculate the unabsorbed X-ray fluxes in the band 0.3--8 keV, by assuming a single power-law model with a photon index of 1.7 and fixed values of $N_{\rm H}$ to corresponding GCs (see Table~\ref{tab:parameters}). If a GC has multiple archival CXO observations, we apply the one with longest exposure time. We note that, since two transient LMXBs have been discovered in the cluster NGC 6440 by {\it Chandra} \citep[see e.g.,][]{intZand2001,Heinke2010}, we need to choose the observation (Obs ID: 947) that was not influenced by the two transients to obtain reasonable X-ray upper limits for the MSPs therein. In addition, if no X-ray counts are detected in the circular region for a MSP, we place the limiting X-ray luminosity of the corresponding GC (see Table~\ref{tab:parameters}) as its upper limit. 

\section{Analysis and Results}
 
\subsection{New X-ray spectral fits in this work}

We used {\sc sherpa},  {\sc ciao}'s modeling and fitting tool, to perform the spectral fits. We assumed fixed hydrogen column densities ($N_{\rm H}$) 
towards corresponding GCs and applied the {\tt xstbabs} model \citep{Wilms2000} to calculate X-ray absorption by the interstellar medium, using {\it wilm} abundances \citep{Wilms2000} and \citet{Verner1996} cross-sections. The $N_{\rm H}$ was estimated based on the known interstellar reddening $E(B-V)$ \citep[][2010 edition]{Harris1996}, and the correlation between $N_{\rm H}$ and the optical extinction $A_V$ \citep{Bahramian2015}, while we adopted $A_V/E(B-V)=3.1$ \citep{Cardelli1989}. We applied the WSTAT statistic \citep{Cash1979} in {\sc sherpa} for estimating the uncertainties of fitting parameters and testing the goodness-of-fit.\footnote{See also \url{https://heasarc.gsfc.nasa.gov/xanadu/xspec/manual/node312.html}.}

\subsubsection{\OC A and B (J1326$-$4728A and B)}

The hydrogen column density towards \OC was fixed at $N_{\rm H}=1.05 \times 10^{21}$ cm$^{-2}$.
Also we binned the spectrum of J1326$-$4728A to contain at least one count per bin, while the data of J1326$-$4728B were binned with two counts per bin.
We considered three spectral models for the X-ray emission from MSPs: blackbody (BB); power-law (PL); and neutron star hydrogen atmosphere \citep[NSATMOS;][]{Heinke2006a} models. We first used the {\tt xsbbody} model to fit BB spectra, and the normalization (with little affect on fitting) was fixed to reduce a free parameter and increase the degrees of freedom. For the NSATMOS spectra, we applied the {\tt xsnsatmos} model, and the NS mass and radius were fixed to 1.4 M$_{\sun}$ and 10 km, respectively. Plus, the distance was frozen to 5.2 kpc, 
our assumed distance to \OC \citep[][2010 revision]{Harris1996}. 
The normalization of the {\tt xsnsatmos} model, physically indicating the fraction of the NS surface emitting, has little influence on other fitting parameters, and therefore we fixed it as well to reduce an unrelated free parameter. We applied {\tt xscflux} model to fit the unabsorbed X-ray fluxes in the band of 0.3--8 keV for BB and NSATMOS spectra ({\tt xscflux} is a convolution model in XSPEC that is usually used for a robust calculation of unabsorbed flux of other model components). As a result, the free parameters were temperature and flux for both BB and NSATMOS models. The PL spectra were fitted using {\tt xspegpwrlw} model, with free parameters of photon index $\Gamma$ and normalization, given that the normalization is in fact the X-ray flux from the source. We applied Q-values to indicate the fitting goodness, which is a measure of probability that the simulated spectra would have a larger reduced statistic value than the observed one, if the assumed model is true and the best-fit parameters are the true parameters. We 
consider Q-values larger than 0.05 as acceptable. The spectral fits of all the three models for J1326$-$4728 A and B are listed in Table~\ref{tab:spec_fits}. 

The spectrum of J1326$-$4728A is well fitted by either a BB model or an NSATMOS model, with effective temperatures of $(2.3 \pm 1.2) \times 10^6$ K and $(1.3 \pm 0.6) \times 10^6$ K, respectively. Also, the fitted unabsorbed fluxes of these two models are consistent with each other. 
A PL spectral model for J1326$-$4728A is 
a slightly worse fit, 
as the best-fit Q-value is 0.06. 
More importantly, 
the fitted photon index is $3.5 \pm 1.1$, which is 
highly unlikely for 
 non-thermal emission from MSPs \citep[typically $\Gamma \sim 1-2$ for MSPs, e.g.][]{Bogdanov2006,Bogdanov2011}. The fitted high photon index 
 implies a soft spectrum 
 typically seen from  
 blackbody-like thermal emission of MSPs \citep[e.g.][]{Zhao2021}. Furthermore, J1326$-$4728A is an isolated MSP \citep{Dai2020} and faint in X-rays ($L_X \sim 2 \times 10^{30}$ erg s$^{-1}$ in 0.3--8 keV band), 
 consistent with a thermal model.

\begin{figure*}
    \centering
    \includegraphics[width=\textwidth]{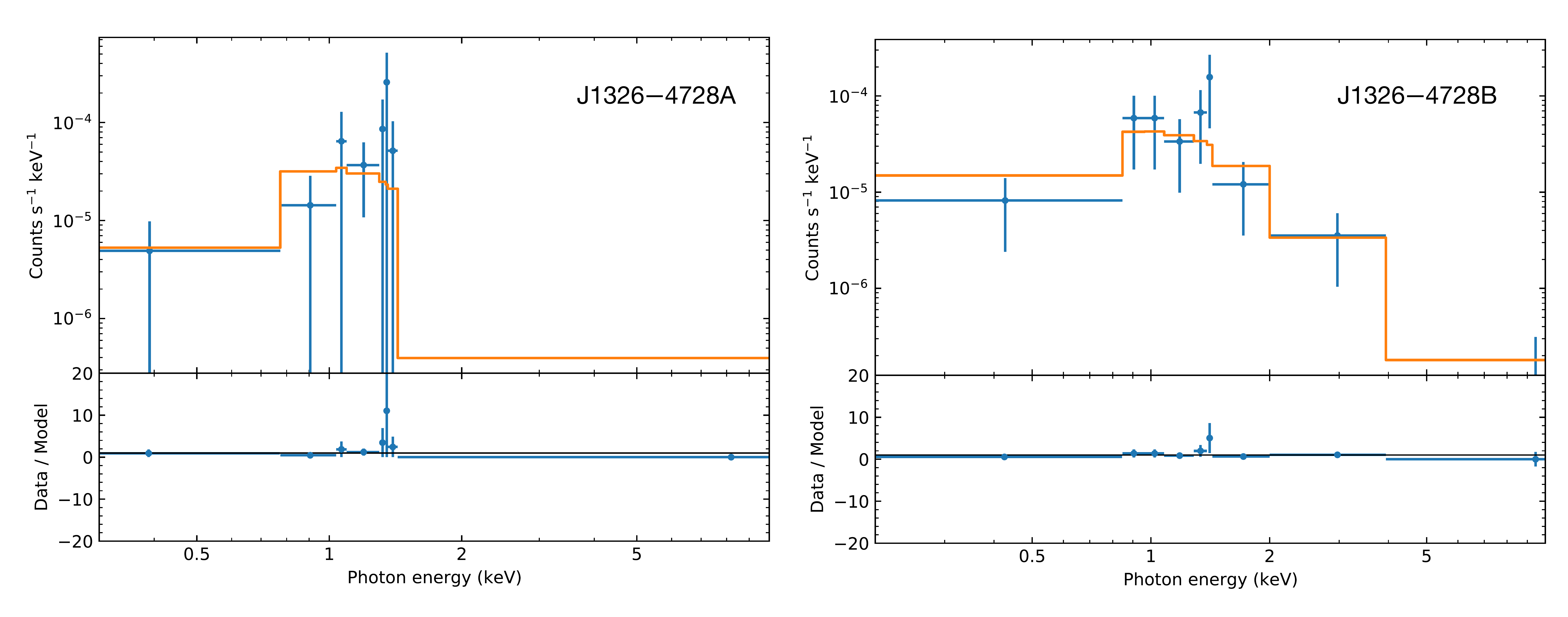}
    \caption{X-ray spectra and best fits of two MSPs in \OC, J1326$-$4728A ({\it left}, fit with an absorbed blackbody) and J1326$-$4728B ({\it right}, fit with an absorbed power-law), respectively, in energy range of 0.3-8 keV. The data of J1326$-$4728A are binned with 1 count/bin, while the data of J1326$-$4728B are binned with 2 count/bin. Both spectra are fitted via the WSTAT statistic. }
    \label{fig:MSP_spec}
\end{figure*}

J1326$-$4728B is an eclipsing black widow pulsar, with a companion star of mass 0.016 M$_{\sun}$ \citep{Dai2020}. Its X-ray spectrum can be well described by a pure PL model, with the fitted photon index $\Gamma = 2.6 \pm 0.5$ and  unabsorbed luminosity $L_X = (1.1 \pm 0.3) \times 10^{31}$ erg s$^{-1}$. It is also likely that the X-ray spectrum is dominated by either a BB or a NSATMOS model, with fitted effective temperatures of $(3.5 \pm 1.2) \times 10^6$ K and $(2.5 \pm 0.6) \times 10^6$ K, respectively, with similar fit quality 'goodness'. Although we cannot rule out any spectral models based on Q-values, we empirically prefer a PL model for the spectrum of J1326$-$4728B, given that most observed eclipsing spider pulsars emit a bulk of non-thermal X-rays, like J0023$-$7203 J and W in 47 Tuc \citep{Bogdanov2006}, 
and given the relatively high $L_X$ of J1326$-$4728B, like these other eclipsing systems. 
However, we need to note that a few nearby MSPs are found with X-ray spectra that can be well described by multiple thermal components plus a PL component \citep[see, e.g.][]{Bogdanov2009,Bogdanov2013}. X-ray spectra of such MSPs with low counts could therefore mimic a single PL model. Though the PL model is preferred for the X-ray spectrum of J1326$-$4728B, the nature of its X-ray emission is still ambiguous.
We also tested combined spectral models, i.e. BB+PL and NSATMOS+PL, but these models did not provide better fits, and hence we only show the fitting results of one-component models here. Figure~\ref{fig:MSP_spec} shows the X-ray spectra and the preferred fits of J1326$-$4728A (left panel) and J1326$-$4728B (right panel).

\begin{table}
    \caption{Spectral fits for J1326$-$4728 A and B.}
    \centering
    \begin{tabular}{lccc}
    \hline
      & \multicolumn{3}{c}{J1326$-$4728A} \\
      \hline
      Spectral Model & BB & PL & NSATMOS  \\
      {$kT_{\rm BB}$/$\Gamma$/$\log T_{\rm eff}$}$^a$ & $0.2 \pm 0.1$ & $3.5 \pm 1.1$ & $6.1 \pm 0.2$ \\
      Reduced Stat. & 1.57 & 2.05 & 1.67 \\
      Q-value & 0.15 & 0.06 & 0.12 \\
      {$F_X$(0.3--8 keV)}$^b$ & $0.6 \pm 0.3 $ & $1.4 \pm 0.9$ & $0.7 \pm 0.3$ \\
      \hline
      \hline
       & \multicolumn{3}{c}{J1326$-$4728B} \\
      \hline
      Spectral Model & BB & PL & NSATMOS  \\
      {$kT_{\rm BB}$/$\Gamma$/$\log T_{\rm eff}$}$^a$ & $0.3 \pm 0.1$ & $2.6 \pm 0.5$ & $6.4 \pm 0.1$ \\
      Reduced Stat. & 0.79 & 1.03 & 0.76 \\
      Q-value & 0.60 & 0.41  & 0.62 \\
      {$F_X$(0.3--8 keV)}$^b$ & $1.8 \pm 0.3 $ & $3.3 \pm 0.9$ & $1.9 \pm 0.5$ \\
      \hline
    \multicolumn{4}{p{0.8\linewidth}}{{\it Notes}: $N_{\rm H}$ was fixed for all the fits at the value to \OC of $1.05 \times 10^{21}$ cm$^{-2}$.} \\
    \multicolumn{4}{p{0.8\linewidth}}{$^a$ $kT_{\rm BB}$: blackbody temperature in units of keV; $\Gamma$: photon index; $\log T_{\rm eff}$: unredshifted effective temperature of the NS surface in units of log Kelvin. } \\
    \multicolumn{4}{l}{$^b$ Unabsorbed flux in units of $10^{-15}$ erg cm$^{-2}$ s$^{-1}$.} \\
    \end{tabular}
    \label{tab:spec_fits}
\end{table}

\subsubsection{M5C (J1518+0204C)}

\begin{figure}
    \centering
    \includegraphics[width=\linewidth]{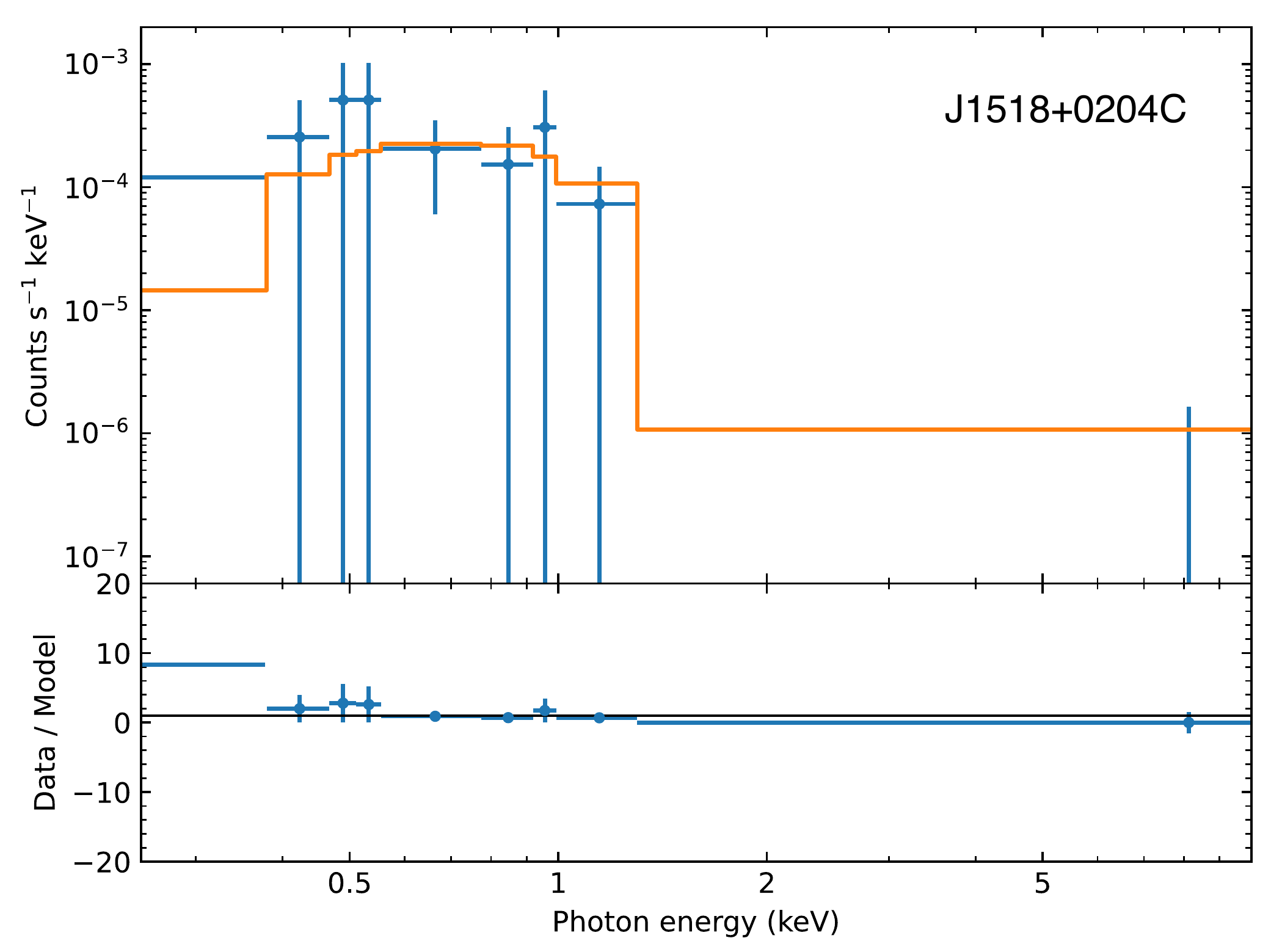}
    \caption{X-ray spectrum and best fit of PSR J1518+0204C in the cluster M5. The data are binned with 1 count/bin and fitted through the WSTAT statistic. The shown spectrum is fitted with an absorbed blackbody model.}
    \label{fig:M5C_spec}
\end{figure}

We fitted the X-ray spectrum of M5C using BB, PL, and NSATMOS models respectively, while the $N_{\rm H}$ towards M5 was fixed at  $2.61 \times 10^{20}$ cm$^{-2}$. Given the limited photon counts from this MSP ($\lesssim$ 10 counts), we needed to manually constrain the normalizations of the BB and NSATMOS models, i.e. the source radius, to obtain 
constraints on other parameters.
Thus we fixed the source radii at 0.3 km for BB model and 1 km for NSATMOS model, respectively, which are commonly observed from GC MSPs \citep[see, e.g.][]{Bogdanov2006}. For the PL model, no additional constraint was required. The fitting results are listed in Table~\ref{tab:spec_fits_M5C}, and the spectrum and the best fit are shown in Figure~\ref{fig:M5C_spec}. 

We found that the X-ray spectrum of M5C is well-described by either a BB or an NSATMOS model, with effective temperatures of $(2.09 \pm 0.02) \times 10^6$ K or $(1.35 \pm 0.09) \times 10^6$ K, respectively. The best-fit unabsorbed fluxes for the BB and NSATMOS models are $1.6^{+0.6}_{-0.5} \times 10^{-15}$ and $1.7^{+0.7}_{-0.5} \times 10^{-15}$ erg cm$^{-2}$ s$^{-1}$, corresponding to X-ray luminosities in the band 0.3--8 keV of $1.1^{+0.4}_{-0.3} \times 10^{31}$ and $1.1^{+0.5}_{-0.3} \times 10^{31}$ \ergs, respectively, at a distance of 7.5 kpc. The fitted photon index in the PL model has $\Gamma = 4.3 \pm 0.8$, also implying a thermally-emitting source with a soft spectrum. 

\begin{table}
    \caption{Spectral fits for J1518$+$0204C in M5.}
    \centering
    \begin{tabular}{lccc}
    \hline
      & \multicolumn{3}{c}{J1518$+$0204C} \\
      \hline
      Spectral Model & BB & PL & NSATMOS  \\
      {$kT_{\rm BB}$/$\Gamma$/$\log T_{\rm eff}$}$^a$ & $0.18 \pm 0.01$ & $4.3 \pm 0.8$ & $6.13 \pm 0.03$ \\
      Reduced Stat. & 1.07 & 0.40 & 1.23 \\
      Q-value & 0.38 & 0.90 & 0.27 \\
      {$F_X$(0.3--8 keV)}$^b$ & $1.6^{+0.6}_{-0.5} $ & $3.6 \pm 1.4$ & $1.7^{+0.7}_{-0.5}$ \\
      \hline
    \multicolumn{4}{p{0.8\linewidth}}{{\it Notes}: $N_{\rm H}$ was fixed for all the fits at the value to M5 of $2.61 \times 10^{20}$ cm$^{-2}$.} \\
    \multicolumn{4}{p{0.8\linewidth}}{$^a$ $kT_{\rm BB}$: blackbody temperature in units of keV; $\Gamma$: photon index; $\log T_{\rm eff}$: unredshifted effective temperature of the NS surface in units of log Kelvin.} \\
    \multicolumn{4}{l}{$^b$ Unabsorbed flux in units of $10^{-15}$ erg cm$^{-2}$ s$^{-1}$.} \\
    \end{tabular}
    \label{tab:spec_fits_M5C}
\end{table}

\subsubsection{M92A (J1717+4308A)}

M92A is a relatively bright X-ray source with 35 counts in its spectrum, and hence we grouped the data with 2 counts per bin to reduce the uncertainties. We fixed the $N_{\rm H}$ towards M92 at $1.74 \times 10^{20}\ {\rm cm}^{-2}$ for the spectral fitting. We first applied three single models, i.e. BB, PL, or NSATMOS (the normalizations for BB and NSATMOS models were thawed), to fit the spectrum, and found that the Q-values for BB and NSATMOS are $\sim$10$^{-3}$ and $\sim$4$\times$10$^{-4}$, respectively. We also checked the fits for BB and NSATMOS models with fixed normalizations and obtained even lower Q-values, and hence these two pure models were ruled out. Although a pure PL model can fairly describe the X-ray spectrum of M92A, we also considered two combined spectral models, PL+BB and PL+NSATMOS, with free normalizations for BB and NSATMOS to fit the data. The fitting results are shown in Table~\ref{tab:spec_fits_M92A}. 

\begin{figure}
    \centering
    \includegraphics[width=\linewidth]{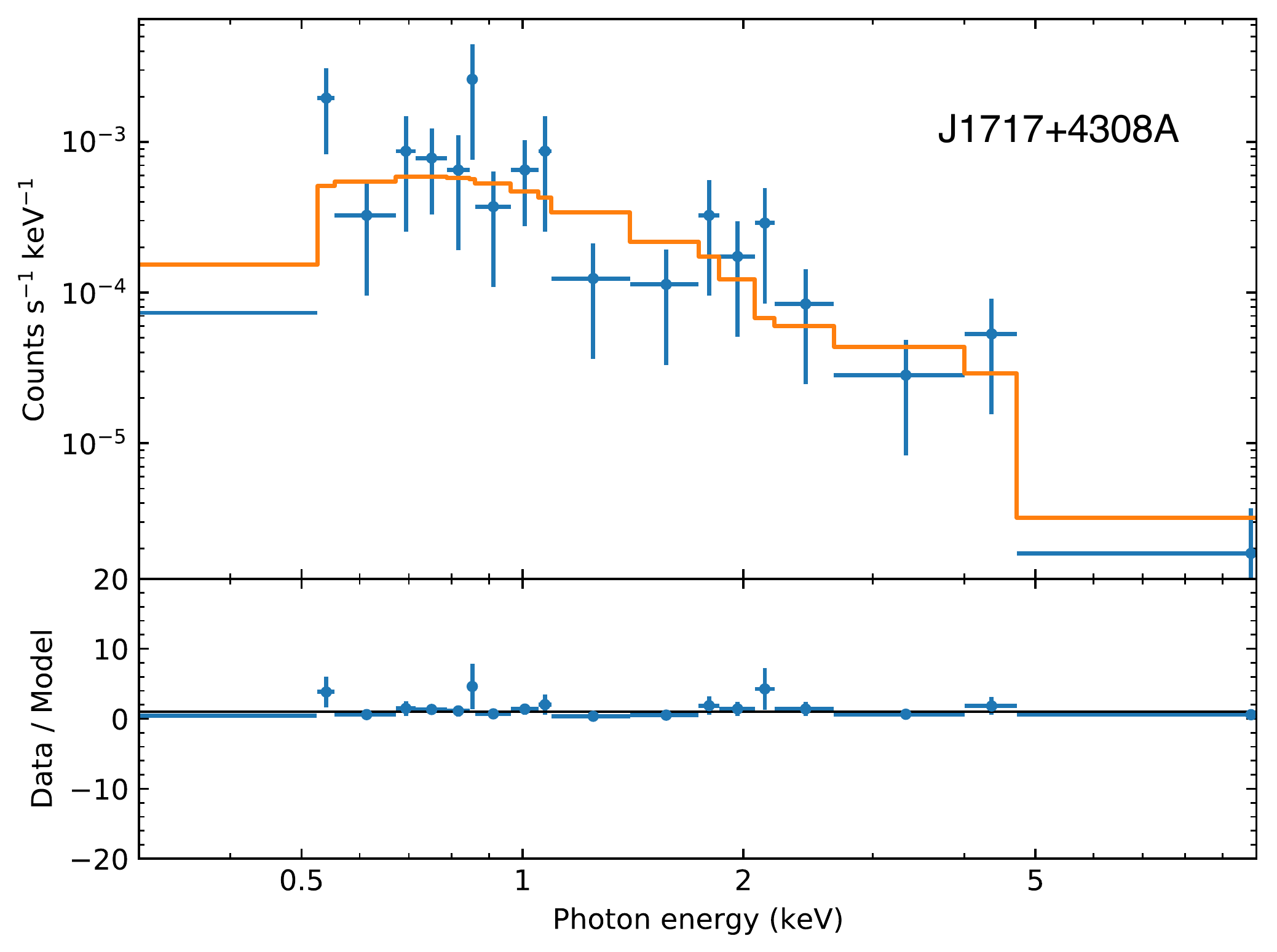}
    \caption{X-ray spectrum and best fit of PSR J1717+4308A in the cluster M92. The data are binned with 2 count/bin and fitted through the WSTAT statistic. The shown spectrum is fitted with a power-law component plus an NSATMOS component.}
    \label{fig:M92A_spec}
\end{figure}

We found that the X-ray spectrum of M92A can be well described by a combined model, either PL+BB or PL+NSATMOS (see Figure~\ref{fig:M92A_spec}), indicating both non-thermal and thermal emission originated from M92A, while the non-thermal component is more dominant. The best-fit photon index is of $\Gamma \sim 1.2$, which is typically observed from redbacks \citep[see][]{Bogdanov2021}. 

The companion star of M92A is most likely a main-sequence star with a median mass of $\sim$0.18 M$_{\sun}$ \citep{Pan2020}, and hence its optical counterpart could be found in the archival observations of {\it Hubble Space Telescope}. \citet{Lu2011} analyzed the optical counterparts of X-ray sources in M92 based on {\rm wavdetect} detections, where the source CX3 in their catalogue  corresponds to M92A. They found nine optical counterparts within the 95\% error circle of CX3, although none of them can yet be confirmed as the counterpart to M92A. Robust identification of the optical counterpart to M92A using its timing position may reveal 
interesting 
properties of its companion star. 

\begin{table}
    \caption{Spectral fits for J1717$+$4308A in M92.}
    \centering
    \begin{tabular}{lccc}
    \hline
      & \multicolumn{3}{c}{J1717$+$4308A} \\
      \hline
      Spectral Model & PL & PL+BB & PL+NSATMOS  \\
      {$\Gamma$}$^a$ & $1.8 \pm 0.3$ & $1.2 \pm 0.6$ & $1.2 \pm 0.5$ \\
      {$kT_{\rm BB}$/$\log T_{\rm eff}$}$^a$ &  $-$ & $0.16 \pm 0.04$ & $6.0 \pm 0.1$ \\
      Reduced Stat. & 1.23 & 1.27 & 1.24 \\
      Q-value & 0.23 & 0.21 & 0.23 \\
      {$F_X$(0.3--8 keV)}$^b$ & $8.8 \pm 1.7 $ & $10.3^{+4.3}_{-3.0}~(7.8^{+2.2}_{-2.2})$ & $10.1^{+2.6}_{-2.6}~(7.6^{+2.1}_{-2.2})$ \\
      \hline
    \multicolumn{4}{p{\linewidth}}{{\it Notes}: $N_{\rm H}$ was fixed for all the fits at the value to M92 of $1.74 \times 10^{20}$ cm$^{-2}$.} \\
    \multicolumn{4}{p{\linewidth}}{$^a$ $kT_{\rm BB}$: blackbody temperature in units of keV; $\Gamma$: photon index; $\log T_{\rm eff}$: unredshifted effective temperature of the NS surface in units of log Kelvin. } \\
    \multicolumn{4}{p{\linewidth}}{$^b$ Unabsorbed flux in units of $10^{-15}$ erg cm$^{-2}$ s$^{-1}$. The values in parentheses represent the flux of the PL component.} \\
    \end{tabular}
    \label{tab:spec_fits_M92A}
\end{table}

\subsubsection{M14A (J1737$-$0314A)}

\begin{figure}
    \centering
    \includegraphics[width=\linewidth]{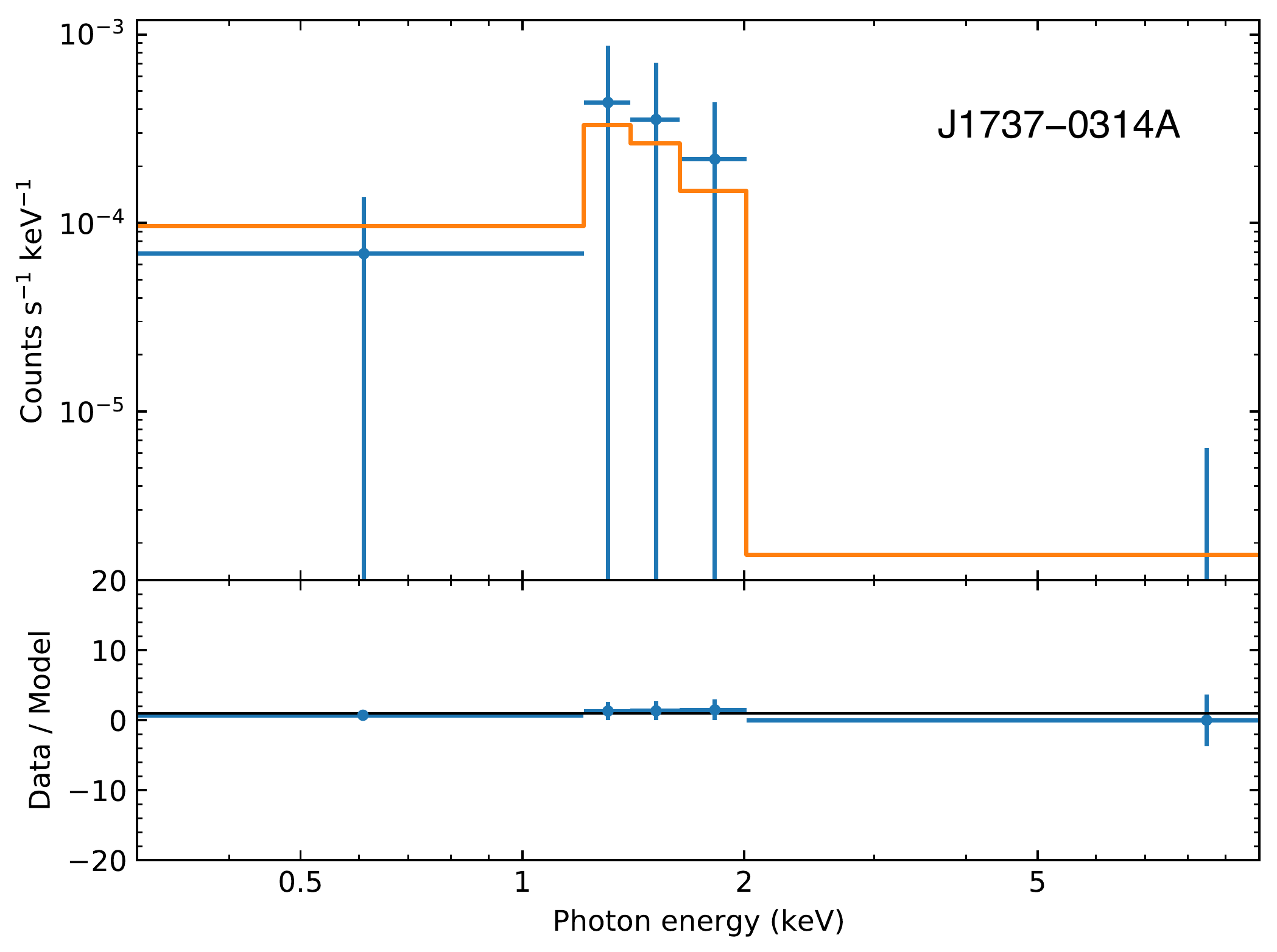}
    \caption{X-ray spectrum and best fit of PSR J1737$-$0314A in the cluster M14. The data are binned with 1 count/bin and fitted through the WSTAT statistic. The shown spectrum is fitted with an absorbed blackbody model.}
    \label{fig:M14A_spec}
\end{figure}

Given that the $N_{\rm H}$ towards M14 is $5.23 \times 10^{21}\ {\rm cm}^{-2}$, the observed X-ray spectrum of M14A is likely affected by the interstellar extinction since soft X-rays ($\lesssim 1$ keV) are largely absorbed (see Figure~\ref{fig:M14A_spec}). We therefore fixed the normalizations in the BB and NSATMOS models with the same values applied for M5C (see above) to get reasonable fits. The fitting results for M14A are listed in Table~\ref{tab:spec_fits_M14A}. 

The X-ray emission generated from M14A is most likely thermal-dominated, given the fitted photon index in the PL model is $\Gamma=3.5\pm0.6$. The spectral fits via a pure BB or a pure NSATMOS model have similar quality. The unabsorbed luminosity (0.3--8 keV) indicated by the BB model is $6.6^{+3.1}_{-2.3} \times 10^{31}$ \ergs at the distance of 9.3 kpc, making it the most X-ray-luminous BW so far.

\begin{table}
    \caption{Spectral fits for J1737$-$0314A in M14.}
    \centering
    \begin{tabular}{lccc}
    \hline
      & \multicolumn{3}{c}{J1737$-$0314A} \\
      \hline
      Spectral Model & BB & PL & NSATMOS  \\
      {$kT_{\rm BB}$/$\Gamma$/$\log T_{\rm eff}$}$^a$ & $0.28 \pm 0.03$ & $3.5 \pm 0.6$ & $6.31 \pm 0.05$ \\
      Reduced Stat. & 0.30 & 0.54 & 0.37 \\
      Q-value & 0.82 & 0.70 & 0.77 \\
      {$F_X$(0.3--8 keV)}$^b$ & $6.4^{+3.0}_{-2.2} $ & $19.9 \pm 10.3$ & $6.8^{+3.7}_{-2.6}$ \\
      \hline
    \multicolumn{4}{p{0.8\linewidth}}{{\it Notes}: $N_{\rm H}$ was fixed for all the fits at the value to M14 of $5.23 \times 10^{21}$ cm$^{-2}$.} \\
    \multicolumn{4}{p{0.8\linewidth}}{$^a$ $kT_{\rm BB}$: blackbody temperature in units of keV; $\Gamma$: photon index; $\log T_{\rm eff}$: unredshifted effective temperature of the NS surface in units of log Kelvin. } \\
    \multicolumn{4}{l}{$^b$ Unabsorbed flux in units of $10^{-15}$ erg cm$^{-2}$ s$^{-1}$.} \\
    \end{tabular}
    \label{tab:spec_fits_M14A}
\end{table}

\subsubsection{NGC 6544A (J1807$-$2459A)}

\begin{figure}
    \centering
    \includegraphics[width=\linewidth]{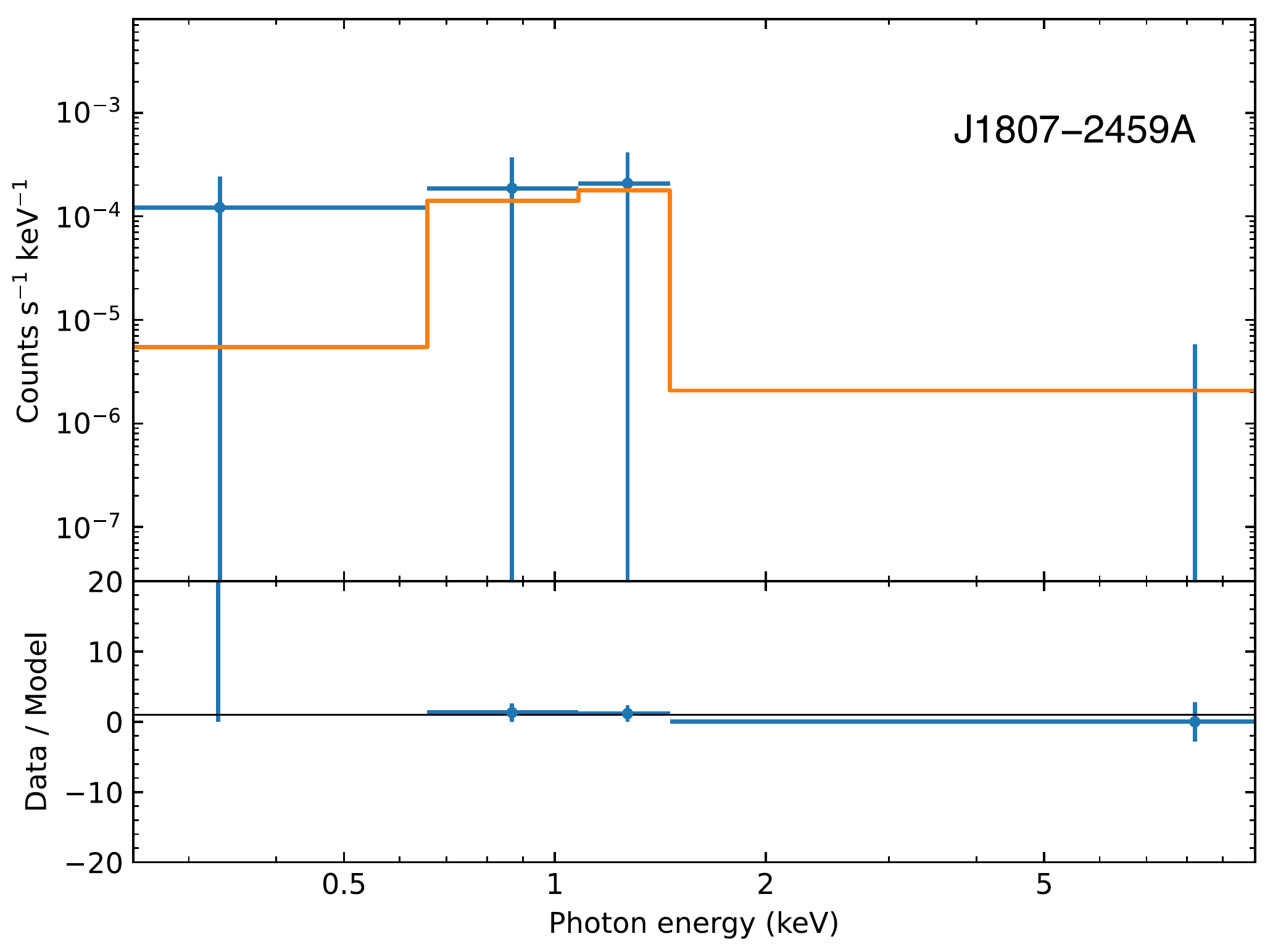}
    \caption{X-ray spectrum and best fit of PSR J1807$-$2459A in the cluster NGC 6544. The data are binned with 1 count/bin and fitted through the WSTAT statistic. The shown spectrum is fitted with an absorbed blackbody model.}
    \label{fig:NGC6544A_spec}
\end{figure}

The observed X-ray spectrum of NGC 6544A is also  significantly impacted by the interstellar extinction towards the cluster (see Figure~\ref{fig:NGC6544A_spec}). Hence, we applied a similar fitting process as used for fitting the spectrum of M14A. By fixing $N_{\rm H} = 6.62 \times 10^{21}$ cm$^{-2}$, we fitted the spectrum using pure BB, PL, and NSATMOS models, respectively, and showed the results in Table~\ref{tab:spec_fits_NGC6544A}. 

NGC 6544A seems also to be a thermally-emitting MSP, with an effective temperature of $2.1 \times 10^6$ K and an unabsorbed luminosity of $9.9^{+5.3}_{-3.9} \times 10^{30}$ \ergs in the BB model. 

\begin{table}
    \caption{Spectral fits for J1807$-$2459A in NGC 6544.}
    \centering
    \begin{tabular}{lccc}
    \hline
      & \multicolumn{3}{c}{J1807$-$2459A} \\
      \hline
      Spectral Model & BB & PL & NSATMOS  \\
      {$kT_{\rm BB}$/$\Gamma$/$\log T_{\rm eff}$}$^a$ & $0.18 \pm 0.02$ & $7.4 \pm 1.8$ & $6.09 \pm 0.06$ \\
      Reduced Stat. & 1.27 & 0.59 & 1.55 \\
      Q-value & 0.28 & 0.55 & 0.20 \\
      {$F_X$(0.3--8 keV)}$^b$ & $9.2^{+4.9}_{-3.6} $ & {$790.9^{+1252.7}_{-790.9}$}$^{\ c}$ & $7.5^{+5.4}_{-3.3}$ \\
      \hline
    \multicolumn{4}{p{0.8\linewidth}}{{\it Notes}: $N_{\rm H}$ was fixed for all the fits at the value to NGC 6544 of $6.62 \times 10^{21}$ cm$^{-2}$.} \\
    \multicolumn{4}{p{0.8\linewidth}}{$^a$ $kT_{\rm BB}$: blackbody temperature in units of keV; $\Gamma$: photon index; $\log T_{\rm eff}$: unredshifted effective temperature of the NS surface in units of log Kelvin. } \\
    \multicolumn{4}{l}{$^b$ Unabsorbed flux in units of $10^{-15}$ erg cm$^{-2}$ s$^{-1}$.} \\
    \multicolumn{4}{l}{$^c$ Model reached lower bound.} \\
    \end{tabular}
    \label{tab:spec_fits_NGC6544A}
\end{table}

\subsection{X-ray sources in M53, NGC 6342, and NGC 6517}
\label{subsec:limLx}

After the procedure of source detection discussed in Section~\ref{subsubsec:src_det}, we found six X-ray sources in M53 and three X-ray sources in each of NGC 6342 and NGC 6517. Basic X-ray information of these detected sources are listed in Table~\ref{tab:X-ray_srcs} (see also Figure~\ref{fig:3_X-ray_images}). 

M53 (or NGC 5024) is the most distant GC studied in this work, at a distance of 17.9 kpc. The interstellar reddening towards M53, however, is relatively low, and hence 
several 
X-ray sources therein can be relatively readily detected. One source is located in the core region, while 
our 
other detected sources are within the half-light radius. The unabsorbed X-ray luminosities (0.3--8 keV) of the six sources are in the range of $3.3 \times 10^{31}$ erg s$^{-1}$ to $7.6 \times 10^{32}$ erg s$^{-1}$, providing an estimated limiting luminosity of $3 \times 10^{31}$ erg s$^{-1}$. 

The source detections of NGC 6342 and NGC 6517 were largely affected by the high ISM absorption and limited exposure time. Particularly, the $N_{\rm H}$ towards NGC 6517 is $\sim 10^{22}$ cm$^{-2}$, among 
the highest $N_{\rm H}$ in this work. The three detected X-ray sources in NGC 6342 have luminosities ranging from $5.1 \times 10^{31}$ erg s$^{-1}$ to $3.8 \times 10^{32}$ erg s$^{-1}$, while the X-ray luminosities of the three sources found in NGC 6517 vary from $2.7 \times 10^{31}$ erg s$^{-1}$ to $2.4 \times 10^{32}$ erg s$^{-1}$ (see Table~\ref{tab:X-ray_srcs}). Hence, the estimated limiting X-ray luminosities of NGC 6342 and NGC 6517 are $5 \times 10^{31}$ erg s$^{-1}$ and $3 \times 10^{31}$ erg s$^{-1}$, respectively. 

For the purpose of this study, we are only interested in the upper and lower limits of the luminosities of X-ray sources in these three GCs. Further studies, like optical/radio identifications of X-ray sources therein, may be available with deeper observations and analysis in the future. 

\begin{table*}
    \centering
    \caption{Basic X-ray properties of catalogue sources in M53, NGC 6342, and NGC 6517. }
    \begin{tabular}{lccccc}
    \hline
    GC	&	Source	&	\multicolumn{2}{c}{Position (J2000)}	&	Counts$^a$	&	$F_X$(0.3--8 keV)$^b$	\\
     & ID & $\alpha$ (hh:mm:ss) & $\delta$ ($\degr$:$\arcmin$:$\arcsec$) & (0.3--8 keV) & ($\times 10^{-15}$ erg cm$^{-2}$ s$^{-1}$) \\
    \hline
    M53	&	1	&	13:12:58.4792 & +18:09:25.378	&	13.9 $\pm$ 3.7	&	4.1	$\pm$	1.1	\\
    M53	&	2	&	13:12:51.7781 & +18:09:33.452	&	34.9 $\pm$ 5.9	&	10.2	$\pm$	1.7	\\
    M53	&	3	&	13:12:52.2771 & +18:09:57.910	&	67.7 $\pm$ 8.2	&	19.8	$\pm$	2.4	\\
    M53	&	4	&	13:12:56.0651 & +18:10:01.569	&	19.7 $\pm$ 4.5	&	5.8	$\pm$	1.3	\\
    M53	&	5	&	13:12:58.5790 & +18:10:29.349	&	2.9	$\pm$ 1.7	&	0.9	$\pm$	0.5	\\
    M53	&	6	&	13:12:55.3503 & +18:10:35.137	&	13.7 $\pm$ 3.7	&	4.0	$\pm$	1.1	\\
    \hline
    NGC 6342	&	1	&	17:21:08.7022 & $-$19:35:54.729	&	20.9 $\pm$ 4.6	&	15.6	$\pm$	3.4	\\
    NGC 6342	&	2	&	17:21:09.7821 & $-$19:35:21.656	&	7.9 $\pm$ 2.8	&	5.9	$\pm$	2.1	\\
    NGC 6342	&	3	&	17:21:09.6572 & $-$19:35:16.323	&	58.9 $\pm$ 7.7	&	43.9	$\pm$	5.7	\\
    \hline
    NGC 6517	&	1	&	18:01:50.6799 & $-$8:57:33.665	&	25.1 $\pm$ 5.1	&	17.8	$\pm$	3.6	\\
    NGC 6517	&	2	&	18:01:51.4020 & $-$8:57:25.036	&	4.8 $\pm$ 2.2	&	3.4	$\pm$	1.6	\\
    NGC 6517	&	3	&	18:01:49.7506 & $-$8:58:02.592	&	2.8 $\pm$ 1.7	&	2.0	$\pm$	1.2	\\
    \hline
    \multicolumn{6}{l}{$^a$ Estimated net counts obtained by {\tt dmextract}; errors are set to 1-$\sigma$.} \\
    \multicolumn{6}{l}{$^b$ Unabsorbed fluxes assuming a PL model with a photon index $\Gamma=1.7$; errors are set to 1-$\sigma$.} \\
    \end{tabular}
    \label{tab:X-ray_srcs}
\end{table*}

\subsection{X-ray luminosity function of GC MSPs}

After data collection and analysis, we finally obtain an X-ray catalogue of 175 GC MSPs (68 MSPs with determined X-ray luminosities shown in Table~\ref{tab:MSP_Lx}, and 107 MSPs with upper limits of X-ray luminosities listed in Table~\ref{tab:appendix_table}). MSPs in NGC 1851, NGC 6441, NGC 6624, NGC 6712, and M15 are not included in this catalogue, since we could not obtain reasonable X-ray luminosity constraints for them due to the severe contamination of very bright X-ray sources in these globular clusters. Also, X-ray luminosities for MSPs in NGC 5986 and NGC 6749 are unavailable because of the lack of sensitive ({\it Chandra}) X-ray observations. 

\begin{figure*}
    \centering
    \includegraphics{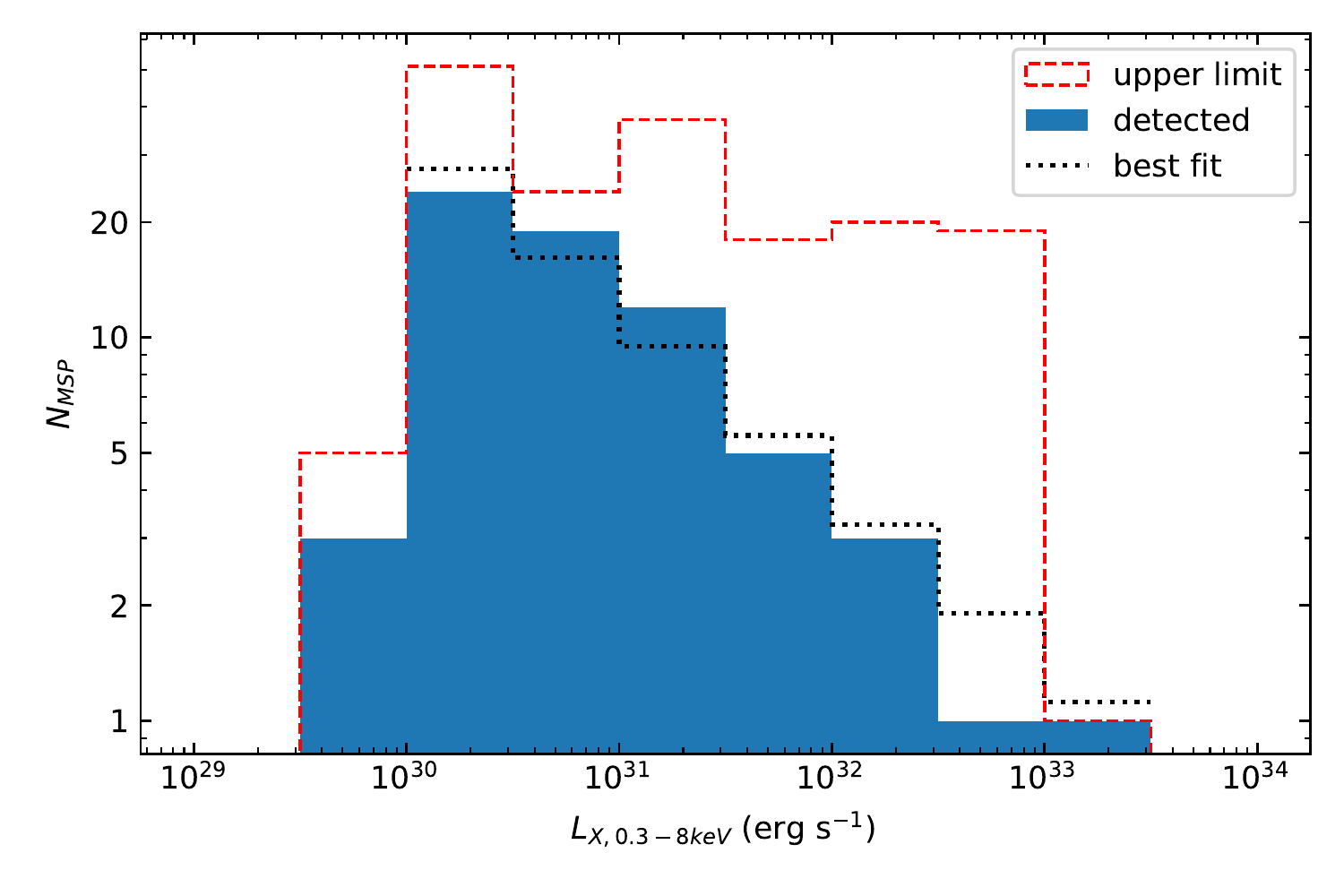}
    \caption{The differential X-ray luminosity distributions of GC MSPs catalogued in this work. The blue histogram shows the distribution of MSPs with determined X-ray luminosities, whereas the red dashed histogram shows the upper limits of the X-ray luminosity distribution. The black dotted histogram shows the best-fit luminosity function for detected MSPs with X-ray luminosities between $10^{30}$ and $\sim 3 \times 10^{33}$ erg s$^{-1}$ (see text for details). }
    \label{fig:LxFunc}
\end{figure*}

Using this catalogue, we are able to empirically investigate the X-ray luminosity function of GC MSPs. Figure~\ref{fig:LxFunc} shows the differential X-ray luminosity distributions for two groups, MSPs with determined X-ray luminosities (blue histogram) and all MSPs in our catalogue including those with upper limits (red dashed histogram), respectively. We found that most detected GC MSPs have X-ray luminosities ranging from $\sim 10^{30}$ to $\sim 3 \times 10^{31}$ erg s$^{-1}$ in the band 0.3--8 keV. And the X-ray luminosity distribution of measured GC MSPs is plausibly a power-law-like pattern, well described by 
$N_{\rm MSP} \propto {L_X}^{-0.46 \pm 0.07}$ 
(see the black dotted histogram in Figure~\ref{fig:LxFunc}; 1-sigma confidence level; though this distribution is likely strongly affected by our not correcting for censorship of the data---that is, by the upper limits).
On the other hand, the upper limits of X-ray luminosities are distributed more evenly between $10^{30}$ and $10^{33}$ erg s$^{-1}$ (likely reflecting the varied X-ray exposures of GCs, principally). 
At the bright end, 
we see the well-known X-ray-bright MSP, B1821$-$24A ($L_X \sim 1.4 \times 10^{33}$ erg s$^{-1}$) in M28, alone.

\begin{figure*}
    \centering
    \includegraphics[width=\textwidth]{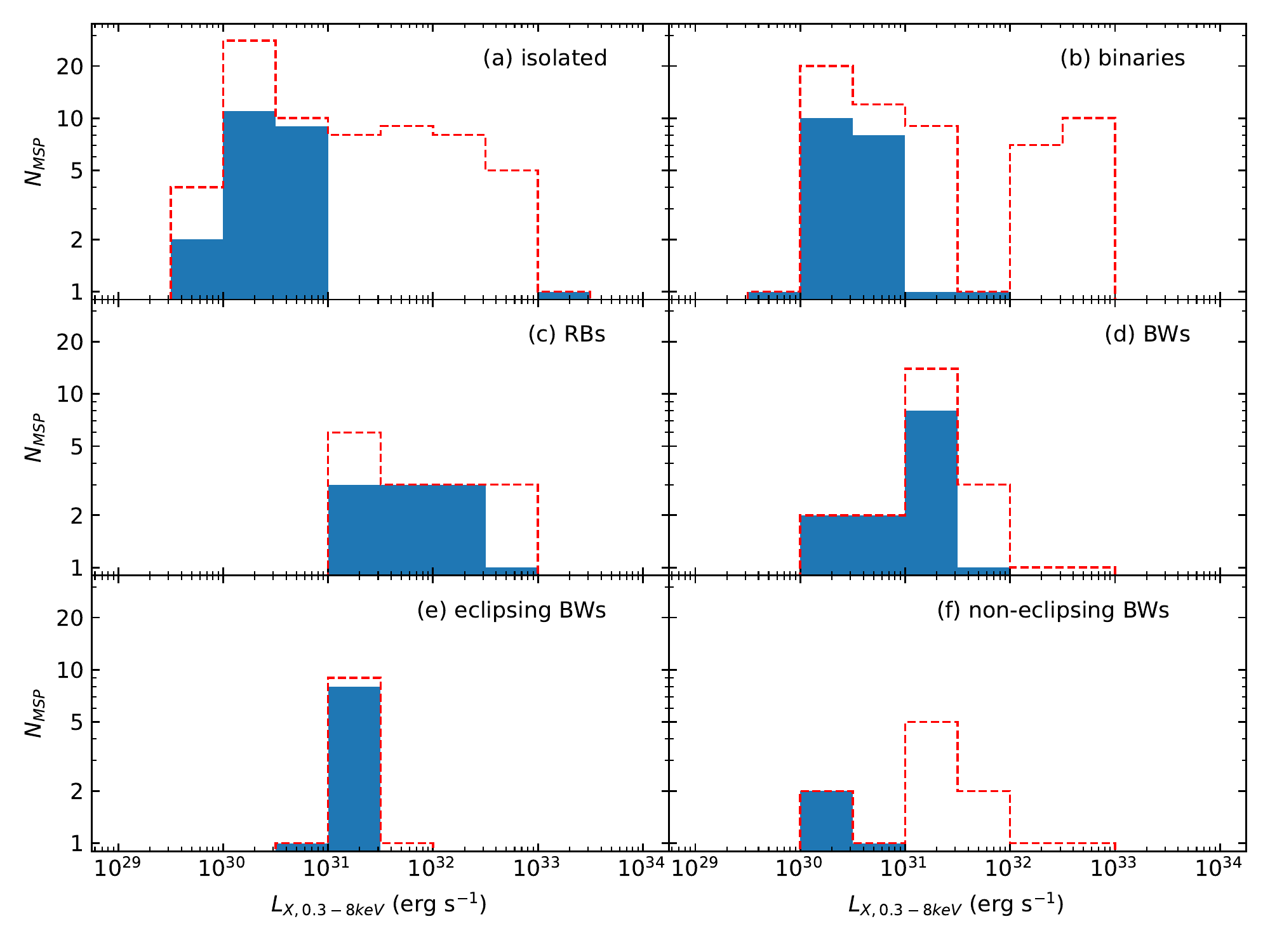}
    \caption{The differential X-ray luminosity functions for (a) isolated MSPs, (b) MSPs in binaries excluding confirmed spider MSPs, (c) redbacks, (d) black widows, (e) eclipsing black widows, and (f) non-eclipsing black widows, respectively, in GCs. The blue histograms show the distributions of MSPs with determined X-ray luminosities, while the red dashed histograms show the upper limits of the X-ray luminosity distributions.}
    \label{fig:LF_sep}
\end{figure*}

We also investigate the X-ray luminosity functions for different MSP groups separately (Figure~\ref{fig:LF_sep}). It is found that all the isolated MSPs in GCs, except M28A, are relatively faint, with X-ray luminosities lower than $10^{30}$ \ergs (see Figure~\ref{fig:LF_sep}a). Most of the MSPs in binary systems, excluding  confirmed spider pulsars, are also faint in X-rays and of similar luminosity distribution with isolated MSPs (Figure~\ref{fig:LF_sep}b). It can interpreted as the same origin of X-ray emission from those MSP systems, that is thermal X-rays generated from hotspots near the NS magnetic poles. 
It is noticeable that all the detected RBs are among the brighter GC MSP population, with X-ray luminosities of 
$1.7\times10^{31}$--$3.4\times10^{32}$
erg s$^{-1}$ (Figure~\ref{fig:LF_sep}c), 
implying that non-thermal X-ray emission produced by intra-binary shocks 
dominates over 
thermal emission from the NS surface. More intriguingly, the X-ray luminosities of detected eclipsing BWs are 
between $L_X$(0.3--8 keV) $=$ $7.0\times10^{30}$ and $2.0\times10^{31}$ erg s$^{-1}$ 
, while the confirmed non-eclipsing BWs are 
almost an order of magnitude 
fainter ($1.5\times10^{30}$ and $3\times10^{30}$ erg s$^{-1}$) than the eclipsing BWs, except one at $9.9\times10^{30}$ \ergs (Figure~\ref{fig:LF_sep}e and f). (Again, additional observations are needed to confirm if J1737$-$0314A is eclipsing, so we did not count it as either an eclipsing or a non-eclipsing BW; see also \ref{subsubsec:M14}).
This observed difference could suggest the population of non-eclipsing black widows are basically the same as the eclipsing black widows except for inclination (as suggested by \citealt{Freire2005b}, noting the lower mass functions of the non-eclipsing systems). In this case, the difference in $L_X$ is also due to inclination (following the models of the intra-binary shocks that indicate the synchrotron X-rays may be beamed in the plane of the binary to some extent).
Alternatively, the population of non-eclipsing black widows are actually different from the eclipsing black widows. These systems may have similar inclinations, but lower-mass companions, that do not produce as strong winds (see e.g. \citealt{Bailes11,Kaplan18}).

However, we need to note some observational biases that may significantly impact the X-ray luminosity distributions of GC MSPs. Given that few GCs have limiting X-ray luminosities lower than $\sim 10^{30}$ erg s$^{-1}$ (see Table~\ref{tab:parameters}), many very faint X-ray MSPs ($L_X < 10^{30}$ erg s$^{-1}$) are not detectable. For instance, 
a few nearby MSPs in the Galactic field have been found to show very faint X-ray luminosities, down to $\sim 1 \times 10^{29}$ \ergs \citep[e.g. PSR J1400$-$1431,][]{Swiggum2017}. Also, 
most MSPs in Terzan 5 remain undetectable in X-rays due to the high absorption towards this GC, even though it has been observed 
extensively by CXO \citep[see e.g.][]{Bogdanov2021}. Furthermore, the upper limits of X-ray luminosities for some MSPs may be  quite conservative. 
Particularly, 
we simply placed one upper limit for 
all the MSPs without timing solutions 
in each GC, defined as the X-ray luminosity of the most luminous unidentified X-ray source in that GC. 
Therefore, we note that the upper limit of X-ray luminosity distribution does not necessarily reflect the estimated number of MSPs in each luminosity range, as they could be orders of magnitude dimmer than the placed upper limits. 

\begin{center}
\begin{table*}
    \caption{Determined X-ray luminosities of GC MSPs}
    \begin{tabular}{lcccccc}
    \hline
    Pulsar Name	&	GC Name	&	Period	&	Type$^a$	&	Spectral Model	&	$L_X$(0.3--8 keV)$^b$	&	Ref. of $L_X$	\\
     & & (ms) & & & ($\times 10^{30}$ erg s$^{-1}$) &  \\
    \hline
    J0023$-$7204C	&	47 Tuc	&	5.76	&	I	&	BB	& $	1.7	^{+	0.3	}_{-	0.6	} $ &	1	\\
    J0024$-$7204D	&	47 Tuc	&	5.36	&	I	&	BB	& $	3.3	^{+	0.4	}_{-	0.8	} $ &	1	\\
    J0024$-$7205E	&	47 Tuc	&	3.54	&	B	&	BB	& $	5.0	^{+	0.6	}_{-	0.9	} $ &	1	\\
    J0024$-$7204F  	&	47 Tuc	&	2.62	&	I	&	BB	& $	2.3	^{+	1.0	}_{-	0.4	} $ &	2	\\
    J0024$-$7204H	&	47 Tuc	&	3.21	&	B	&	BB	& $	3.2	^{+	0.3	}_{-	0.8	} $ &	1	\\
    J0023$-$7203J	&	47 Tuc	&	2.10	&	eBW	&	BB+PL	& $	11.6	^{+	2.3	}_{-	3.7	} $ &	1	\\
    J0024$-$7204L	&	47 Tuc	&	4.35	&	I	&	BB	& $	8.6	^{+	0.8	}_{-	1.0	} $ &	1	\\
    J0023$-$7205M	&	47 Tuc	&	3.68	&	I	&	BB	& $	2.4	^{+	0.3	}_{-	0.7	} $ &	1	\\
    J0024$-$7204N	&	47 Tuc	&	3.05	&	I	&	BB	& $	2.4	^{+	0.4	}_{-	0.7	} $ &	1	\\
    J0024$-$7204O	&	47 Tuc	&	2.64	&	eBW	&	BB+PL	& $	10.8	^{+	3.5	}_{-	2.1	} $ &	1	\\
    J0024$-$7204Q	&	47 Tuc	&	4.03	&	B	&	BB	& $	2.4	^{+	0.3	}_{-	0.7	} $ &	1	\\
    J0024$-$7204R	&	47 Tuc	&	3.48	&	eBW	&	BB	& $	7.0	^{+	0.4	}_{-	1.4	} $ &	1	\\
    J0024$-$7204S	&	47 Tuc	&	2.83	&	B	&	BB	& $	4.2	^{+	0.7	}_{-	0.7	} $ &	2	\\
    J0024$-$7204T	&	47 Tuc	&	7.59	&	B	&	BB	& $	1.5	^{+	0.3	}_{-	0.6	} $ &	1	\\
    J0024$-$7203U	&	47 Tuc	&	4.34	&	B	&	BB	& $	3.2	^{+	0.3	}_{-	0.8	} $ &	1	\\
    J0024$-$7204W  	&	47 Tuc	&	2.35	&	eRB	&	BB+PL	& $	26.4	^{+	1.0	}_{-	6.3	} $ &	1	\\
    J0024$-$7201X	&	47 Tuc	&	4.77	&	B	&  BB &	$2.2^{+0.6}_{-0.6}$	&	3	\\
    J0024$-$7204Y  	&	47 Tuc	&	2.20	&	B	&	BB	& $	2.5	^{+	0.2	}_{-	0.7	} $ &	1	\\
    J0024$-$7205Z  	&	47 Tuc	&	4.55	&	I	&	BB	& $	3.5	^{+	0.5	}_{-	0.5	} $ &	2	\\
    J0024$-$7205aa 	&	47 Tuc	&	3.69	&	I	&	BB	& $	0.9	^{+	0.4	}_{-	0.3	} $ &	2	\\
    J0024$-$7204ab 	&	47 Tuc	&	3.70	&	I	&	BB	& $	2.0	^{+	0.4	}_{-	0.5	} $ &	2	\\
    J1326$-$4728A	&	\OC	&	4.11	&	I	&	BB	& $	1.9	^{+	1.0	}_{-	1.0	} $ &	4	\\
    J1326$-$4728B	&	\OC	&	4.79	&	eBW	&	PL	& $	10.7	^{+	3.0	}_{-	3.0	} $ &	4	\\
    J1518$+$0204C & M5 & 2.48 & eBW & BB & $10.8^{+4.0}_{-3.8}$ & 4 \\
    B1620$-$26	&	M4	&	11.08	&	O	&	BB	& $	3.0	^{+	0.7	}_{-	0.7	} $ &	5	\\
    J1641$+$3627B	&	M13	&	3.53	&	B	&	PL	& $	9.2	^{+	4.6	}_{-	4.6	} $ &	6	\\
    J1641$+$3627C	&	M13	&	3.72	&	I	&	BB	& $	3.9	^{+	1.3	}_{-	0.7	} $ &	6	\\
    J1641$+$3627D	&	M13	&	3.12	&	B	&	BB	& $	5.9	^{+	1.3	}_{-	0.7	} $ &	6	\\
    J1641$+$3627E	&	M13	&	2.49	&	eBW	&	PL	& $	12.5	^{+	4.6	}_{-	4.6	} $ &	6	\\
    J1641$+$3627F	&	M13	&	3.00	&	B	&	BB	& $	7.9	^{+	1.3	}_{-	0.7	} $ &	6	\\
    J1701$-$3006B	&	M62	&	3.59	&	eRB	&	PL	& $	101.0	^{+	25.9	}_{-	21.0	} $ &	7	\\
    J1701$-$3006C	&	M62	&	7.61	&	B	&	PL	& $	59.0	^{+	12.0	}_{-	12.0	} $ &	7	\\
    J1717$+$4308A & M92 & 3.16 & eRB & BB+PL & $83.3^{+21.4}_{-21.4}$ & 4 \\
    J1737$-$0314A & M14 & 1.98 & BW$^{\star}$ & BB & $66.2^{+31.1}_{-22.8}$ & 4 \\
    J1740$-$5340A	&	NGC 6397	&	3.65	&	eRB	&	PL	& $	22.2	^{+	2.6	}_{-	2.5	} $ &	7	\\
    J1740$-$5340B	&	NGC 6397	&	5.79	&	eRB	&	PL	& $	67.0	^{+	0.5	}_{-	0.1	} $ &	8, 9	\\
    J1748$-$2446A	&	Terzan 5	&	11.56	&	eRB	&	PL	& $	89.9	^{+	27.2	}_{-	27.2	} $ &	10	\\
    J1748$-$2446E	&	Terzan 5	&	2.20	&	B	&	PL	& $	1.9	^{+	0.7	}_{-	0.2	} $ &	10	\\
    J1748$-$2446F	&	Terzan 5	&	5.54	&	I	&	PL	& $	5.3	^{+	1.9	}_{-	1.5	} $ &	10	\\
    J1748$-$2446H	&	Terzan 5	&	4.93	&	I	&	PL	& $	4.4	^{+	1.7	}_{-	1.4	} $ &	10	\\
    J1748$-$2446K	&	Terzan 5	&	2.97	&	I	&	PL	& $	1.3	^{+	0.6	}_{-	0.5	} $ &	10	\\
    J1748$-$2446L	&	Terzan 5	&	2.24	&	I	&	PL	& $	8.3	^{+	2.0	}_{-	1.8	} $ &	10	\\
    J1748$-$2446N  	&	Terzan 5	&	8.67	&	B	&	PL	& $	1.4	^{+	0.6	}_{-	0.4	} $ &	10	\\
    J1748$-$2446O	&	Terzan 5	&	1.68	&	eBW	&	PL	& $	20.3	^{+	3.2	}_{-	3.2	} $ &	10	\\
    J1748$-$2446P	&	Terzan 5	&	1.73	&	eRB	&	PL	& $	335.6	^{+	21.4	}_{-	18.4	} $ &	10	\\
    J1748$-$2446Q        	&	Terzan 5	&	2.81	&	B	&	PL	& $	0.6	^{+	0.5	}_{-	0.4	} $ &	10	\\
    J1748$-$2446V        	&	Terzan 5	&	2.07	&	B	&	PL	& $	21.0	^{+	9.8	}_{-	7.0	} $ &	10	\\
    J1748$-$2446X	&	Terzan 5	&	3.00	&	B	&	PL	& $	3.1	^{+	1.3	}_{-	1.3	} $ &	10	\\
    J1748$-$2446Z	&	Terzan 5	&	2.46	&	B	&	PL	& $	6.8	^{+	1.3	}_{-	1.3	} $ &	10	\\
    J1748$-$2446ad	&	Terzan 5	&	1.40	&	eRB	&	PL	& $	139.1	^{+	16.6	}_{-	12.5	} $ &	10	\\
    J1807$-$2459A & NGC 6544 & 3.06 & BW & BB & $9.9^{+5.3}_{-3.9}$ & 4 \\
    B1821$-$24A	&	M28	&	3.05	&	I	&	PL	& $	1375.7	^{+	47.1	}_{-	32.6	} $ &	11	\\
    J1824$-$2452C	&	M28	&	4.16	&	B	&	PL	& $	2.0	^{+	0.6	}_{-	0.5	} $ &	11	\\
    J1824$-$2452E	&	M28	&	5.42	&	I	&	PL	& $	2.3	^{+	0.7	}_{-	0.6	} $ &	11	\\
    J1824$-$2452F	&	M28	&	2.45	&	I	&	PL	& $	1.4	^{+	0.4	}_{-	0.3	} $ &	11	\\
    J1824$-$2452H	&	M28	&	4.63	&	eRB	&	PL	& $	17.4	^{+	3.3	}_{-	13.0	} $ &	11	\\
    J1824$-$2452I	&	M28	&	3.93	&	eRB	&	PL	& $	220.0	^{+	40.0	}_{-	40.0	} $ &	12	\\
    J1824$-$2452J	&	M28	&	4.04	&	BW	&	PL	& $	1.5	^{+	0.2	}_{-	0.2	} $ &	11	\\
    J1824$-$2452K	&	M28	&	4.46	&	B	&	PL	& $	6.2	^{+	0.9	}_{-	0.9	} $ &	11	\\
    J1836$-$2354A	&	M22	&	3.35	&	BW	&	PL	& $	3.0	^{+	1.6	}_{-	1.0	} $ &	13	\\
    \end{tabular}
    \label{tab:MSP_Lx}
\end{table*}

\begin{table*}
    \begin{tabular}{lcccccc}
    \hline
    Pulsar Name	&	GC Name	&	Period	&	Type$^a$	&	Spectral Model	&	$L_X$(0.3--8 keV)$^b$	&	Ref. of $L_X$	\\
     & & (ms) & & & ($\times 10^{30}$ erg s$^{-1}$) &  \\
    \hline
    J1910$-$5959E	&	NGC 6752	&	4.57	&	I	&	BB	& $	1.0	^{+	1.0	}_{-	0.4	} $ &	14	\\
    J1910$-$5959F	&	NGC 6752	&	8.49	&	I	& PL &	$4.0^{+1.0}_{-0.7}$	&	15	\\
    J1953$+$1846A	&	M71	&	4.89	&	eBW	&	PL	& $	12.0	^{+	1.9	}_{-	1.9	} $ &	16	\\
    J2140$-$2310A        	&	M30	&	11.02	&	eBW	&	PL	& $	10.9	^{+	4.2	}_{-	3.6	} $ &	17	\\
    J1911$-$5958A	&	NGC 6752	&	3.27	&	B	&	BB	& $	2.9	^{+	1.3	}_{-	1.0	} $ &	14	\\
    J1910$-$5959B	&	NGC 6752	&	8.36	&	I	&	BB	& $	1.3	^{+	0.6	}_{-	0.4	} $ &	14	\\
    J1911$-$6000C	&	NGC 6752	&	5.28	&	I	&	BB	& $	3.2	^{+	0.8	}_{-	0.6	} $ &	14	\\
    J1910$-$5959D	&	NGC 6752	&	9.04	&	I	&	BB	& $	3.8	^{+	0.8	}_{-	0.6	} $ &	14	\\
    \hline
    \end{tabular}
    \contcaption{\\ {\it Notes}: periods and types of MSPs were obained from Paulo Freire’s GC Pulsar Catalog.\footnote{\url{http://www.naic.edu/~pfreire/GCpsr.html}} \\
    $^a$ Types of MSP systems; I: isolated; B: binary; BW: black widow; RB: redback; O: others; e: eclipsing. \\
    $^b$ Unabsorbed X-ray luminosities at the distances to corresponding GCs. \\
    $^{\star}$ J1737$-$0314A in M14 might be an eclipsing BW. Confirmation is needed. See text for more information. \\
    References: 1) \citet{Bogdanov2006}; 
    2) \citet{Bhattacharya2017}; 
    3) \citet{Ridolfi2016};
    4) this work; 
    5) \citet{Pavlov2007}; 
    6) \citet{Zhao2021}; 
    7) \citet{Oh2020}; 
    8) \citet{Bogdanov2010}; 
    9) \citet{PichardoMarcano2021}; 
    10) \citet{Bogdanov2021}; 
    11) \citet{Bogdanov2011}; 
    12) \citet{Linares2014}; 
    13) \citet{Amato2019}; 
    14) \citet{Forestell2014};
    15) Cohn et al. (2021, submitted); 
    16) \citet{Elsner2008}; 
    17) \citet{Zhao2020b}.}
    \label{tab:MSP_ Lx_continued}
\end{table*}
\end{center}

\subsection{Number of MSPs versus stellar encounter rate}

We re-examined the correlation between the number of MSPs and stellar encounter rate (\SE) for GCs in this work (see Table~\ref{tab:parameters}). It is well established that the number of X-ray binaries in a GC has a strong correlation with the GC stellar encounter rate \citep[see e.g.][]{Pooley2003,Heinke2003,Bahramian2013}. One would naturally 
assume 
a correlation between the number of MSPs and \SE in a GC, since MSPs are 
offspring of LMXBs. However, the difficulty to establish such a correlation is determining the total number of radio MSPs harboured in a GC. \citet{Bagchi2011} used sophisticated Monte Carlo simulations with various radio luminosity functions and models to calculate the population of radio MSPs in 10 GCs. 
They claimed that they did not find strong evidence that the number of MSPs correlates with \SE under either model. 
However, 
\citet{Bahramian2013} 
produced a more sophisticated calculation of \SE for these GCs 
and adopted the estimates of MSP populations from \citet[model 1]{Bagchi2011}, finding a significant correlation between the number of MSPs and \SE in a GC via their statistical tests. 

\begin{figure*}
    \centering
    \includegraphics{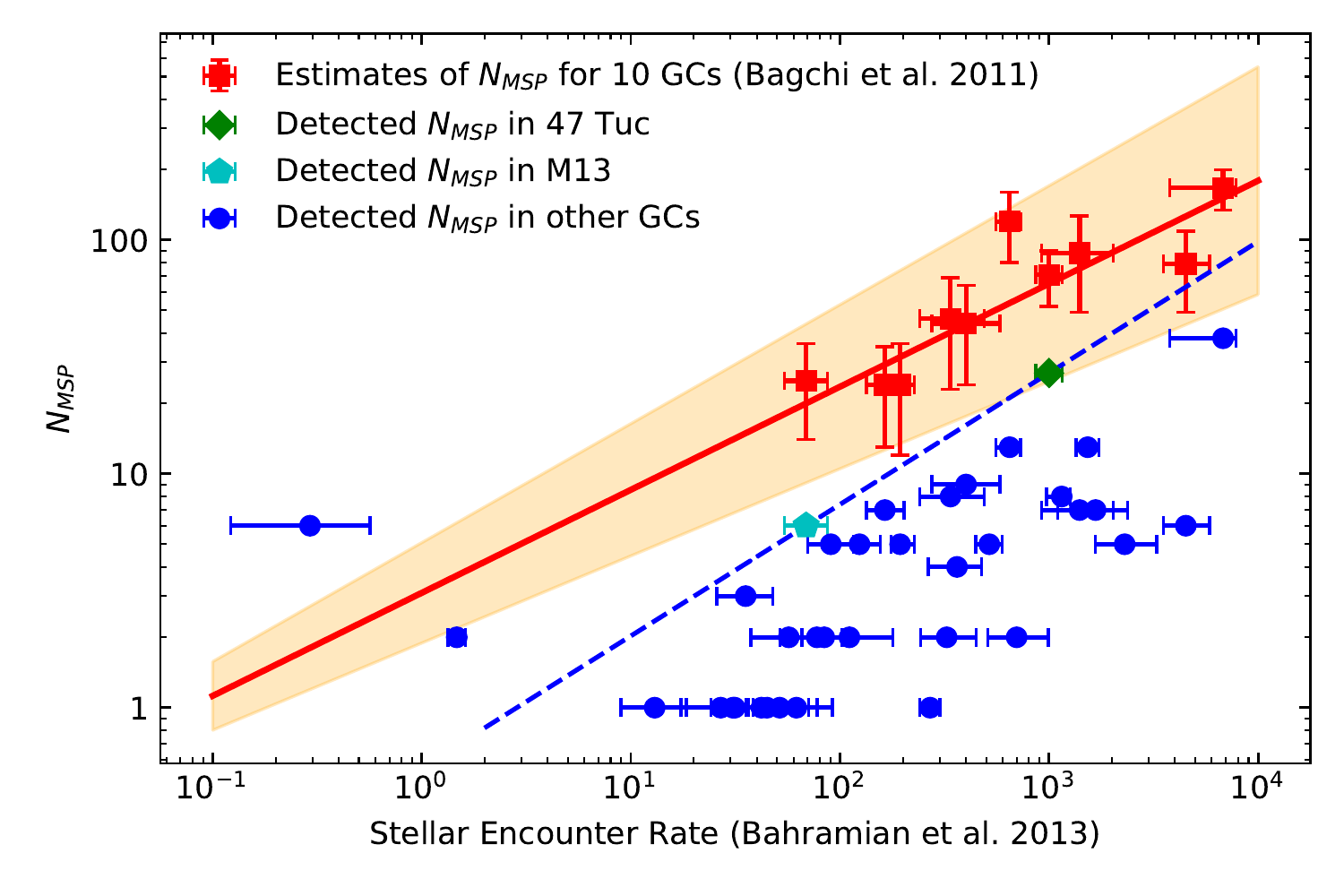}
    \caption{The number of MSPs in a globular cluster versus the stellar encounter rate. The blue dots show the numbers of currently known radio MSPs in GCs, 
    while the red squares show the estimated MSP population for 10 GCs calculated by \citet[model 1]{Bagchi2011}. The red solid line shows the best-fit power-law model using the data points from \citet{Bagchi2011}, and the yellow shade represents the 1-sigma interval. The blue dashed line shows the power-law model fitted to the data points of 47 Tuc and M13 (see text for details). All the stellar encounter rates of GCs and their corresponding errors were extracted from \citet{Bahramian2013}. }
    \label{fig:N-Gamma}
\end{figure*}

Here, we assume both the \SE values from \citet{Bahramian2013} and the calculations of MSP population for 10 GCs (we also adopt the results of model 1 here; column 2 in Table~\ref{tab:estimates}) from \citet{Bagchi2011} are correct, and fit a power-law function through the data points, as 
standard
for the correlation between the number of XRBs and \SE.
We note that due to the lack of data points, Markov Chain Monte Carlo (MCMC) method could not provide robust results, and hence we applied orthogonal distance regression method for fitting. 
The best-fit curve is $\log (N_{\rm MSP})=0.44 \log ($\SE$) + 0.49$, with 1-sigma errors on the slope and intercept of $\pm 0.07$ and $\pm 0.21$, respectively (red solid line in Figure~\ref{fig:N-Gamma}). Based on this correlation (assuming it is true) and the fitted function, we can then roughly estimate the total number of MSPs for each GC for a given \SE (column 3 in Table~\ref{tab:estimates}). 
We calculate 
approximately 1460 MSPs in total in the 36 GCs in our study, which we consider as a conservative upper estimate. 
We consider this as a conservative upper estimate, because \citet{Bagchi2011} also produce two other estimates with smaller predicted numbers: roughly half those of Model 1. In addition, \citet{Heinke2005} place an upper limit on the MSP population of 47 Tuc of $<$60 at 95\% confidence, compared to Model 1's 71$\pm$19. Alternative estimates from gamma-rays \citep{Abdo2010} and diffuse radio flux \citep{McConnell04} also are well below Bagchi et al.'s Model 1  numbers --- 33$\pm15$ and $\lesssim$ 30, respectively.
Some MSPs may not point their radio beams towards us (though \citealt{Lyne88} illustrate that beam width varies inversely with spin period, so this fraction is likely to be small). 
However, the limits from X-ray and gamma-ray studies above do not depend on the radio beaming fraction, as the gamma-rays are produced high in the magnetosphere with a very large beaming fraction \citep{Venter09}, and X-rays are emitted from hot spots on the neutron star surface, which are gravitationally lensed so as to be visible to virtually all observers \citep{Pechenick83}. 

We also plotted the number of currently known radio MSPs versus \SE for the GCs in this work (blue dots in Figure~\ref{fig:N-Gamma}; see also Table~\ref{tab:parameters}). Furthermore, we made an aggressive estimate of the lower bound on the total population of GC MSPs. Unlike the conservative upper estimate, we consider two GCs, 47 Tuc and M13, as ``well-determined'' GCs, for which the currently found MSPs in these two GCs are all (or nearly all) the MSPs therein. The choice of 47 Tuc is based on comprehensive and extensive studies of this cluster. Particularly, \citet{Heinke2005} suggested a total number of $\sim 25$ MSPs in 47 Tuc by comparing X-ray colors, luminosities, variability, etc., of detected MSPs to those of unidentified sources using deep CXO observations. (See also the compatible estimates using gamma-rays and diffuse radio flux, mentioned above.)  To date, 27 MSPs have been found in 47 Tuc (four without timing positions), and 
this aggressive estimate assumes that
this is indeed close to the total number of MSPs therein. M13 was recently observed by \citet{Wang2020} using the Five-hundred-metre Aperture Spherical radio Telescope (FAST). A new MSP (J1641+3627F) was discovered in their observations, making a total of 6 MSPs found in this cluster. Since the sensitivity of their radio observations reached down to a flux density of 0.4 $\mu$Jy to a candidate with a signal to noise ratio of 7 \citep{Wang2020}, corresponding to a pseudo radio luminosity of $\sim$0.02 mJy kpc$^2$ at 1.4 GHz, below which no MSPs in GCs have been detected 
(most GC MSPs have radio specific luminosities of $\gtrsim 1$ mJy kpc$^2$ at 1.4 GHz),  we can confidently assume all or nearly all the MSPs in M13 have been found. We therefore fitted a power-law model across these two data points, finding a correlation of $\log (N_{\rm MSP})=0.56 \log ($\SE$) - 0.26$ (blue dashed line in Figure~\ref{fig:N-Gamma}).  We normalized the MSP populations in other GCs to this model, assuming an MSP population proportional to their stellar encounter rates, and thus  obtained 
a rough lower bound (henceforth our "aggressive" estimate) of the 
 MSP population for them (see column 4 in Table~\ref{tab:estimates}). 

It is noticeable that there is one cluster, Terzan 1 (leftmost blue dot in Figure~\ref{fig:N-Gamma}), that falls significantly outside both our aggressive and conservative model predictions. Since Terzan 1 is a so-called `core-collapsed' and heavily obscured cluster and its structural parameters are poorly measured, the calculation of \SE depending on core values, such as core density and core radius, is generally not considered reliable  \citep[see][]{Cackett2006,Bahramian2013}. An alternative explanation of the discrepancy between the number of MSPs (and also of XRBs) and \SE in Terzan 1 is that most stars in Terzan 1 have been stripped due to Galactic tides, leaving a core unusually rich in binaries \citep{deMarchi99}.
This is particularly plausible for Terzan 1, considering that it is located very close to the Galactic centre  \citep{Cackett2006}. 

\begin{table}
    \centering
    \caption{Estimates of MSP population in GCs}
    \label{tab:estimates}
    \begin{tabular}{lccccc}
    \hline
    GC Name	& 	N$_{\rm calc}^a$	&	N$_{\rm cons}^b$	&	N$_{\rm aggr}^c$	&	N$_{\rm unid}^d$	&	Ref.	\\
    \hline
    47 Tuc (NGC 104)	& $	71	\pm	19	$ &	 $\sim 65 $	 & $	27^e	$ &	0	&	1	\\
    NGC 1851	& $		-		$ &	$\sim	78	$ & $ \sim	34	$ &	$-$	&		\\
    M53 (NGC 5024)	& $		-		$ &	$\sim	15	$ & $ \sim	4	$ &	5	&	2	\\
    \OC (NGC 5139)	& $		-		$ &	$\sim	23	$ & $ \sim	7	$ &	7	&	3	\\
    M3 (NGC 5272)	& $	24	\pm	12	$ &	$\sim	32	$ & $ \sim	11	$ &	2	&	4	\\
    M5 (NGC 5904)       	& $	24	\pm	11	$ &	$\sim	29	$ & $ \sim	10	$ &	1	&	5	\\
    NGC 5986	& $		-		$ &	$\sim	19	$ & $ \sim	6	$ &	$-$	&		\\
    M4 (NGC 6121)	& $		-		$ &	$\sim	13	$ & $ \sim	4	$ &	0	&	6	\\
    M13 (NGC 6205)	& $	25	\pm	11	$ &	$\sim	20	$ & $ 	6^e	$ &	3	&	5,7	\\
    M12 (NGC 6218)	& $		-		$ &	$\sim	10	$ & $ \sim	2	$ &	0	&	8	\\
    M10 (NGC 6254)	& $		-		$ &	$\sim	14	$ & $ \sim	4	$ &	0	&	5	\\
    M62 (NGC 6266)	& $		-		$ &	$\sim	82	$ & $ \sim	36	$ &	4	&	9	\\
    M92 (NGC 6341)	& $		-		$ &	$\sim	37	$ & $ \sim	13	$ &	5	&	10	\\
    NGC 6342	& $		-		$ &	$\sim	17	$ & $ \sim	5	$ &	2	&	2	\\
    Terzan 1	& $		-		$ &	$\sim	2	$ & $ \sim	0	$ &	0	&	11	\\
    M14 (NGC 6402)	& $		-		$ &	$\sim	26	$ & $ \sim	8	$ &	4	&	5	\\
    NGC 6397	& $		-		$ &	$\sim	22	$ & $ \sim	7	$ &	0	&	12	\\
    Terzan 5	& $	167	\pm	33	$ &	$\sim	151	$ & $ \sim	79	$ &	26	&	13	\\
    NGC 6440	& $	88	\pm	39	$ &	$\sim	75	$ & $ \sim	33	$ &	4	&	14	\\
    NGC 6441	& $		-		$ &	$\sim	94	$ & $ \sim	43	$ &	$-$	&		\\
    NGC 6517	& $	46	\pm	23	$ &	$\sim	40	$ & $ \sim	15	$ &	1	&	2	\\
    NGC 6522	& $		-		$ &	$\sim	42	$ & $ \sim	15	$ &	1	&	5	\\
    NGC 6539	& $		-		$ &	$\sim	16	$ & $ \sim	5	$ &	4	&	5	\\
    NGC 6544	& $		-		$ &	$\sim	25	$ & $ \sim	8	$ &	1	&	5	\\
    NGC 6624	& $		-		$ &	$\sim	69	$ & $ \sim	29	$ &	$-$	&		\\
    M28 (NGC 6626)	& $	120	\pm	40	$ &	$\sim	54	$ & $ \sim	21	$ &	7	&	5	\\
    NGC 6652	& $		-		$ &	$\sim	56	$ & $ \sim	22	$ &	3	&	15	\\
    M22 (NGC 6656)	& $		-		$ &	$\sim	21	$ & $ \sim	6	$ &	1	&	5	\\
    NGC 6712	& $		-		$ &	$\sim	14	$ & $ \sim	4	$ &	$-$	&		\\
    NGC 6749	& $		-		$ &	$\sim	18	$ & $ \sim	5	$ &	$-$	&		\\
    NGC 6752	& $	44	\pm	20	$ &	$\sim	43	$ & $ \sim	16	$ &	0	&	16	\\
    NGC 6760	& $		-		$ &	$\sim	18	$ & $ \sim	5	$ &	0	&	5	\\
    M71 (NGC 6838)	& $		-		$ &	$\sim	4	$ & $ \sim	1	$ &	0	&	17,18	\\
    M15 (NGC 7078)	& $	79	\pm	30	$ &	$\sim	126	$ & $ \sim	63	$ &	$-$	&		\\
    M2 (NGC 7089)	& $		-		$ &	$\sim	49	$ & $ \sim	19	$ &	6	&	5	\\
    M30 (NGC 7099)	& $		-		$ &	$\sim	40	$ & $ \sim	14	$ &	0	&	19	\\
    \hline
    Total & $-$ & $\sim 1457$ & $\sim 587$ & $87$ & \\
    \hline
    \multicolumn{6}{p{\linewidth}}{{\it Notes}: $^a$ Calculations of MSP population for 10 GCs by \citet{Bagchi2011}} \\
    \multicolumn{6}{l}{$^b$ Conservative estimates of GC MSP population} \\
    \multicolumn{6}{l}{$^c$ Aggressive estimates of GC MSP population} \\
    \multicolumn{6}{l}{$^d$ Unidentified X-ray sources with $L_X > 10^{32}$ erg s$^{-1}$ (0.3--8 keV)} \\
    \multicolumn{6}{l}{$^e$ Selected as normalization} \\
    \multicolumn{6}{p{\linewidth}}{Reference of N$_{\rm unid}$: (1) \citet{Heinke2005}; (2) this work; (3) \citet{Henleywillis2018}; (4) \citet{Zhao2019}; (5) \citet{Bahramian2020}; (6) \citet{Bassa2004}; (7) \citet{Servillat2011}; (8) \citet{Lu2009}; (9) \citet{Oh2020}; (10) \citet{Lu2011}; (11) \citet{Cackett2006}; (12) \citet{Bogdanov2010}; (13) \citet{Heinke2006b}; (14) \citet{Pooley2002}; (15) \citet{Stacey2012}; (16) \citet{Forestell2014}; (17) \citet{Elsner2008}; (18) \citet{Huang2010}; (19) \citet{Zhao2020b}} \\
    \end{tabular}
\end{table}

\section{Discussion}

The radio MSPs in \OC were not discovered until a more advanced receiver, an ultra-wide-bandwidth low-frequency receiver \citep{Hobbs2020}, was installed and used on the Parkes radio telescope. Before that, however, some studies in other bands had hinted at the existence of MSPs in \OC. For example, \citet{Abdo2010} suggested a total of $19 \pm 9$ MSPs harboured in \OC based on  gamma-ray 
detection of the cluster 
by the {\it Fermi} Large Area Telescope (LAT). A deep X-ray survey of \OC by \citet{Henleywillis2018} also implied the presence of MSPs, given that tens of unidentified sources have similar X-ray colours with the MSPs detected in 47 Tuc. Intriguingly, the X-ray counterpart to one of the MSPs in \OC (J1326$-$4728B; analyzed in this work) has been detected in previous X-ray studies \citep[][source ID 13d in their tables]{Haggard2009,Henleywillis2018}. 

However, it is possible to nominate unidentified sources detected in other bands (e.g. X-rays and gamma-rays) as radio MSP candidates based on their observational properties. In fact, a large number of radio MSPs have recently been discovered by targeting LAT unassociated sources \citep[see e.g.][]{Ray2012}. Moreover, dedicated analysis with X-ray, gamma-ray and optical observations also can lead to discoveries of new MSPs. For instance,  \citet{Bogdanov2010} suggested an X-ray source in the cluster NGC 6397 (source ID U18 in their work) as a strong MSP candidate, given its similar X-ray and optical properties to those of the known MSP, PSR J1740$-$5340, in the cluster. Later, \citet{Zhao2020a} found the radio counterpart to U18 using 
the Australia Telescope Compact Array, while \citet{PichardoMarcano2021} reported the optical modulation of the companion star to U18, and both of their studies provided strong evidence that U18 is a ``hidden'' redback MSP. 
The radio pulsations from this source have been detected recently by the Parkes radio telescope (Lei Zhang et al. 2021, in prep.), verifying it as a redback MSP (PSR J1740$-$5340B). The reason of previous non-detection of radio pulsations can be interpreted as  scattering of the radio pulsations by wind from the companion  \citep{Zhao2020a}. 

Considering the fact that there might be a group of hidden MSPs observed in X-rays but without radio confirmation, like PSRs J1326$-$4728B and J1740$-$5340B, it is also interesting to investigate the faint unidentified X-ray sources, especially those with $L_X > 10^{32}$ erg s$^{-1}$, where the non-thermal X-ray emission dominates. We 
count 
all the unidentified X-ray sources with $L_X > 10^{32}$ erg s$^{-1}$ (0.3--8 keV) in our studied GCs 
in Table~\ref{tab:estimates} (column 5). We found a total of 87 unidentified X-ray sources with $L_X > 10^{32}$ erg s$^{-1}$ in 29 GCs, 
while we did not find any unidentified sources with $L_X > 10^{33}$ erg s$^{-1}$ in those GCs. However, we note a special case, the X-ray source B in NGC 6652 \citep[see][]{Heinke2001,Stacey2012}, which is suggested as a transitional MSP (tMSP) in a recent work by \citet{Paduano2021}. It is considered to be 
in an accretion-powered state currently, with an average X-ray luminosity of $\sim 1.8 \times 10^{34}$ erg s$^{-1}$. However, since we cannot completely confirm the nature of this source until its rotation-powered MSP state is detected, we just treat it as an unidentified source with $L_X > 10^{33}$ erg s$^{-1}$ in this work. 
Terzan 5 contains a large number of unidentified X-ray sources, which we attribute to its high stellar encounter rate (the highest among known GCs, \citealt{Bahramian2013}), and the large interstellar extinction towards the cluster \citep{Massari12}, which makes identification of optical counterparts extremely difficult \citep[cf.][]{Testa12,Ferraro15}; a general search for optical/infrared counterparts of Terzan 5 X-ray sources has not yet been conducted.
Another two interesting GCs are M14 and M2, where the MSPs therein were discovered recently by FAST \citep{Pan2021b}. All the detected MSPs in these two clusters are found in binary systems, while one BW and two eclipsing RB MSPs were found in M14. Since it is common that eclipsing RBs have X-ray luminosities of $L_X \gtrsim 10^{32}$ erg s$^{-1}$ (see Table~\ref{tab:MSP_Lx}), the X-ray counterparts to the two newly found RBs might be included in those unidentified sources. However, due to the lack of deep X-ray observations and radio timing solutions of these MSPs, 
investigation of their X-ray properties is not yet possible.

Our results 
allow an estimate of the X-ray detectable 
MSP population in the Galactic centre. 
An excess of gamma-rays, peaking at $\sim$ 2 GeV, has been found towards the Galactic centre \citep[also known as the Galactic Centre Excess, or GCE, see e.g.,][]{Hooper2011,Gordon2013,Ajello2016,Daylan2016}, 
the origin of which remains unclear.
Some studies suggested that the excess is generated from dark matter annihilation \citep[e.g.][]{Hooper2011,Daylan2016,Ackermann2017,diMauro2021}, whereas other groups claimed that a population of unresolved MSPs in the Galactic bulge produces the observed gamma-ray excess \citep[e.g.,][]{Abazajian2012,Brandt2015,Gonthier2018,Macias2018}. If 
a large number of MSPs 
are present in the Galactic centre, one can also expect they are emitting X-rays. 
While {\it Fermi} LAT has a relatively poor angular resolution of ($\sim$1 degree),\footnote{\url{https://fermi.gsfc.nasa.gov/science/instruments/table1-1.html}} {\it Chandra}'s  high angular resolution of 0.5 arcsec\footnote{\url{https://asc.harvard.edu/proposer/POG/}} allows the possible detection of X-ray counterparts to MSPs as point sources in the Galactic centre. 
However, the large interstellar extinction towards the Galactic centre absorbs nearly all X-ray emission below 2 keV, rendering faint MSPs producing only soft blackbody-like emission (the majority of MSPs) undetectable.
Only a few X-ray-bright MSPs with substantial magnetospheric (such as PSR B1821$-$24A in M28) or shock-powered (redbacks) X-ray emission might be detected around the Galactic centre by {\it Chandra}, or even {\it XMM-Newton} (with its larger point-spread function of $\sim$10"). 

Using our X-ray census of GC MSPs and our estimates of the total population of MSPs in those GCs, we are able to 
estimate the population of easily detectable X-ray MSPs ($L_X > 10^{33}$ erg s$^{-1}$ in the band 0.3--8 keV) in the Galactic bulge. We first adopt the prediction by \citet{Gonthier2018}, who suggested a total of $\sim$11,000 MSPs in the Galactic bulge are needed to produce the GCE. 
The number of MSPs required to explain the GCE has been estimated at 10,000--20,000 \citep{Yuan2014}, 2,000--14,000 \citep{Cholis2015}, $\sim$40,000 \citep{Ploeg2017}, or $\sim$10,000 \citep{Gonthier2018}, and consequently our estimation may also vary from other 
predictions of the number of MSPs. 
To calculate a lower limit of detectable X-ray MSPs in the Galactic center, we assume PSR B1821$-$24A is the only MSP with an X-ray luminosity more than $10^{33}$ erg s$^{-1}$ among 1500 MSPs (the conservative estimate in Table~\ref{tab:estimates}), and the MSP population in the Galactic bulge keeps the same proportion. 
Allowing for small-number statistics \citep{Gehrels1986} gives an estimate of $>$1 MSP above $10^{33}$ erg/s for 8,700 MSPs (a 1$\sigma$ lower limit), and thus predicts of order 1 
MSP with $L_X>10^{33}$ erg/s in the Galactic bulge. 
On the other hand, if we assume the upper limit of the number of MSPs with $L_X > 10^{33}$ erg s$^{-1}$ (2; 
considering NGC 6652B) and 
take the ``aggressive'' lower estimate of 590 MSPs in these GCs, 
then we obtain an upper limit (1$\sigma$) of easily detectable MSPs of 1 MSP above $10^{33}$ erg/s for 126 MSPs, and thus predict of order 86 such easily detectable MSPs in the Galactic Centre.
Performing the same calculation for  $L_X > 10^{32}$ erg/s (still very achievable with {\it Chandra} in the Galactic Centre) gives a predicted range of 20 to 910 X-ray detectable MSPs in the Galactic Centre.
These predictions are consistent with those of \citet{Berteaud2020}, which used an alternative method of inferring X-ray luminosity functions from gamma-ray luminosity functions. 
Careful study of X-ray sources in the Galactic Bulge may be able to identify plausible MSP candidates, or rule out such an MSP candidate population, which would favour a dark matter interpretation for the GCE.
We will look into available {\it Chandra} and {\it XMM-Newton} observations towards the Galactic Centre to identify candidate MSPs therein, in future works. 



\begin{table}
    \centering\
    \caption{Estimates of the number of X-ray MSPs in the bulge}
    \begin{tabular}{lccc}
    \hline
        $L^a_X$(erg s$^{-1}$) & N$^b_{\rm det}$ & N$^c_{\rm upp}$ & N$^d_{\rm est}$ \\
    \hline
        $> 10^{32}$ & 5 & 41 & $21-908$ \\
       $> 3 \times 10^{32}$ & 2 & 20 &  $5-480$ \\
        $> 10^{33}$ & 1 & 2 & $1.2-86$ \\
    \hline
    \multicolumn{4}{l}{$^a$Unabsorbed luminosities in 0.3--8 keV} \\
    \multicolumn{4}{l}{$^b$Number of detected MSPs in GCs} \\ 
    \multicolumn{4}{l}{$^c$Upper limit of the number of MSPs in GCs} \\
    \multicolumn{4}{l}{$^d$Estimated number of MSPs in the bulge} \\
    \end{tabular}
    \label{tab:N_MSP_in_GC}
\end{table}

\section{Conclusions}

In this work, we compiled X-ray luminosities of MSPs in GCs, including new X-ray analysis.
We analysed the X-ray spectra of two MSPs (PSR J1326$-$4728A and J1326$-$4728B) in the cluster \OC. The unabsorbed X-ray luminosities in the band 0.3--8 keV of these two MSPs are $\sim 2 \times 10^{30}$ erg s$^{-1}$ and $\sim 1 \times 10^{31}$ erg s$^{-1}$, respectively. The X-ray spectrum of PSR J1326$-$4728A is well described by either a BB or a NSATMOS model, indicating thermal X-ray emission from the neutron star surface. The spectrum of PSR J1326$-$4728B is well-fit by a PL model, with a photon index of $2.6 \pm 0.5$. Its spectrum reflects the bulk of non-thermal X-ray emission from the MSP, which is commonly observed from eclipsing spider pulsars and likely produced by intra-binary shocks. 
We also presented new X-ray analyses for PSRs J1518$+$0204C, J1717$+$4308A, J1737$-$0314A, and J1807$-$2459A in GCs M5, M92, M14, and NGC 6544, respectively. The X-ray spectrum of J1717$+$4308A is well described by a composition of a PL model and a BB model, with a photon index of 1.2$\pm$0.6 and an effective temperature of (1.9$\pm$0.5)$\times$10$^6$ K. The unabsorbed luminosity in the band 0.3--8 keV of J1717$+$4308A is (8.3$\pm$2.1)$\times$10$^{31}$ \ergs, consisting of both thermal and non-thermal emission. The spectra of other three newly analyzed MSPs (J1518$+$0204C, J1737$-$0314A, and J1807$-$2459A) are well fitted by a single BB model, with X-ray luminosities (0.3--8 keV) ranging from $\sim$1.1$\times$10$^{31}$ \ergs to 6.6$\times$10$^{31}$ \ergs.
We also 
catalogued 
the X-ray sources in the clusters M53, NGC 6342, and NGC 6517. We found a total of 12 X-ray sources in these three GCs, with X-ray luminosities ranging from $\sim 3 \times 10^{31}$ erg s$^{-1}$ to $\sim 8 \times 10^{32}$ erg s$^{-1}$. 

We presented a comprehensive census of X-ray MSPs in 29 Galactic GCs.
We reported the X-ray luminosities or upper limits for 175 GC MSPs in our catalogue, and normalized the energy band to 0.3--8 keV. We determined X-ray luminosities for 68 GC MSPs and constrained the luminosities for others, except  MSPs in 2 GCs that have no {\it Chandra} observations, and in 5 GCs that are severely contaminated by bright X-ray sources. We investigated the empirical MSP X-ray luminosity function using our catalogue, finding that most detected GC X-ray MSPs have luminosities between $\sim 1 \times 10^{30}$ erg s$^{-1}$ and $\sim 3 \times 10^{31}$ erg s$^{-1}$. The X-ray luminosities for eclipsing spider MSPs are generally higher than other types of MSPs, with $L_X \gtrsim 10^{31}$ erg s$^{-1}$. 

We re-examined the correlation between the number of MSPs and stellar encounter rate in a GC. Using the estimates of numbers of MSPs in several clusters from \citet{Bagchi2011} and the stellar interaction rates of \citet{Bahramian2013}, we found a relation of $\log (N_{\rm MSP})=0.44 \log ($\SE)$ + 0.49$, which we take as an upper limit to the numbers of MSPs in clusters. We also estimated a lower limit using the  numbers of known MSPs in the most well-observed clusters 47 Tuc and M13, and assuming that other clusters follow the same relation of stellar interaction rate and number of MSPs, 
$\log (N_{\rm MSP})=0.56 \log $(\SE)$ - 0.26$. 
We estimated the total number of MSPs in each of the GCs in this work using both fitting relations, and suggested a conservative upper estimate of 
1500 MSPs and an aggressive lower estimate of 
590 
MSPs, respectively, in those 36 GCs.

We empirically estimated the population of detectable MSPs in the Galactic  bulge, assuming the gamma-ray excess is produced by a large number of unresolved MSPs. Based on our census of GC X-ray MSPs, we suggested 
of order 1-90 
MSPs with $L_X > 10^{33}$ erg s$^{-1}$ in the Galactic centre, and of order 20-900 MSPs with $L_X > 10^{32}$ erg s$^{-1}$ there. 
As these sources are likely 
detected in existing archival 
{\it Chandra} and XMM-Newton observations, dedicated searches
may uncover the proposed Galactic Bulge MSP population. 

\section*{Acknowledgements}

We thank Sharon Morsink, Erik Rosolowsky and Gregory Sivakoff for helpful discussions. JZ thanks Zhichen Pan for the updated information of the timing position of M92A. 
COH is supported by NSERC Discovery Grant RGPIN-2016-04602. 
JZ is supported by the China Scholarship Council (CSC). 
This work has made use of data obtained from the Chandra Data Archive and the Chandra Source Catalogue, and software provided by the Chandra X-ray Centre (CXC) in the application packages {\sc ciao}, {\sc sherpa}, {\sc ds9}, and {\sc pimms}. This work has made use of data from the European Space Agency (ESA) mission
{\it Gaia} (\url{https://www.cosmos.esa.int/gaia}), processed by the {\it Gaia}
Data Processing and Analysis Consortium (DPAC,
\url{https://www.cosmos.esa.int/web/gaia/dpac/consortium}). Funding for the DPAC
has been provided by national institutions, in particular the institutions
participating in the {\it Gaia} Multilateral Agreement. 
This research has made use of NASA's Astrophysics Data System Bibliographic Services and arXiv. This research has made use of the VizieR catalogue access tool, CDS, Strasbourg, France (DOI : 10.26093/cds/vizier). The original description of the VizieR service was published in 2000, A\&AS 143, 23

\section*{Data Availability}

The {\it Chandra} data used in this article are available in the Chandra Data Archive (\url{https://cxc.harvard.edu/cda/}) by searching the Observation ID listed in Table~\ref{tab:obs_GCs} and Table~\ref{tab:obs_OC} in the Search and Retrieval interface, ChaSeR (\url{https://cda.harvard.edu/chaser/}). The {\it Gaia} data used in this work are available in the VizieR Information System (\url{https://vizier.cds.unistra.fr/index.gml}).



\bibliographystyle{mnras}
\bibliography{example} 




\appendix

\section{Upper limits of X-ray luminosities} 

We present the upper limits of X-ray luminosities for 107 GC MSPs in Table~\ref{tab:appendix_table}. These MSPs are not eligible for performing specific X-ray spectral analysis, and hence we placed an upper limit of $L_X$ for each of them. The criteria of determining upper limits have been discussed in Section~\ref{sec:data}.

\begin{table*}
    \centering
    \caption{107 GC MSPs with upper limits of X-ray luminosities.}
    \begin{tabular}{lccccc}
\hline
Pulsar Name	&	GC Name	&	Period	&	Type	&	$L_X$(0.3--8 keV)	&	References	\\
 & & (ms) & & {($\times$10$^{30}$ \ergs)} &  \\
\hline
J0024$-$7204G*	&	47 Tuc (NGC 104)	&	4.04	&	I	& {	20	} &	1	\\
J0024$-$7204I*	&	47 Tuc (NGC 104)	&	3.48	&	BW	& {	20	} &	1	\\
J0024$-$7204P	&	47 Tuc (NGC 104)	&	3.64	&	BW	& {	20	} &	2	\\
J0024$-$7204V	&	47 Tuc (NGC 104)	&	4.81	&	eRB	& {	20	} &	2	\\
J0024$-$7204ac	&	47 Tuc (NGC 104)	&	2.74	&	eBW	& {	20	} &	3	\\
J0024$-$7204ad	&	47 Tuc (NGC 104)	&	3.74	&	eRB	& {	20	} &	3	\\
J1312+1810B	&	M53 (NGC 5024)	&	6.24	&	B	& {	760	} &	4	\\
J1312+1810C	&	M53 (NGC 5024)	&	12.53	&	I	& {	760	} &	4	\\
J1312+1810D	&	M53 (NGC 5024)	&	6.07	&	B	& {	760	} &	4	\\
J1326$-$4728C	&	\OC (NGC 5139)	&	6.87	&	I	& {	300	} &	5	\\
J1326$-$4728D	&	\OC (NGC 5139)	&	4.58	&	I	& {	300	} &	5	\\
J1326$-$4728E	&	\OC (NGC 5139)	&	4.21	&	I	& {	300	} &	5	\\
J1342+2822A	&	M3 (NGC 5272)	&	2.55	&	BW	& {	13	} &	6	\\
J1342+2822B*	&	M3 (NGC 5272)	&	2.39	&	B	& {	9	} &	6	\\
J1342+2822C	&	M3 (NGC 5272)	&	2.17	&	I	& {	13	} &	6	\\
J1342+2822D*	&	M3 (NGC 5272)	&	5.44	&	B	& {	13	} &	6	\\
J1342+2822E	&	M3 (NGC 5272)	&	5.47	&	B	& {	13	} &	6	\\
J1342+2822F	&	M3 (NGC 5272)	&	4.4	&	B	& {	13	} &	6	\\
B1516+02A*	&	M5 (NGC 5904)	&	5.55	&	I	& {	2	} &	7	\\
B1516+02B*	&	M5 (NGC 5904)	&	7.95	&	B	& {	2	} &	7	\\
J1518+0204D	&	M5 (NGC 5904)	&	2.99	&	B	& {	160	} &	9	\\
J1518+0204E	&	M5 (NGC 5904)	&	3.18	&	B	& {	160	} &	9	\\
J1518+0204F	&	M5 (NGC 5904)	&	2.65	&	B	& {	160	} &	10	\\
J1518+0204G	&	M5 (NGC 5904)	&	2.75	&	B	& {	160	} &	11	\\
J1641+3627A*	&	M13 (NGC 6205)	&	10.38	&	I	& {	1	} &	12	\\
J1647$-$0156A	&	M12 (NGC 6218)	&	2.36	&	B	& {	190	} &	13	\\
J1657$-$0406A	&	M10 (NGC 6254)	&	4.73	&	-	& {	59	} &	14	\\
J1657$-$0406B	&	M10 (NGC 6254)	&	7.35	&	B	& {	59	} &	14	\\
J1701$-$3006A*	&	M62 (NGC 6266)	&	5.24	&	B	& {	4	} &	15	\\
J1701$-$3006D*	&	M62 (NGC 6266)	&	3.42	&	B	& {	7	} &	16	\\
J1701$-$3006E*	&	M62 (NGC 6266)	&	3.23	&	eBW	& {	56	} &	16	\\
J1701$-$3006F*	&	M62 (NGC 6266)	&	2.29	&	BW	& {	36	} &	16	\\
J1701$-$3006G	&	M62 (NGC 6266)	&	4.61	&	B	& {	100	} &	17	\\
J1721$-$1936B	&	NGC 6342	&	2.57	&	I	& {	380	} &	19	\\
J1735$-$3028B	&	Terzan 1	&	11.14	&	I	& {	85	} &	20	\\
J1735$-$3028C	&	Terzan 1	&	6.04	&	I	& {	85	} &	20	\\
J1735$-$3028D	&	Terzan 1	&	5.39	&	I	& {	85	} &	20	\\
J1735$-$3028E	&	Terzan 1	&	3.08	&	I	& {	85	} &	20	\\
J1735$-$3028F	&	Terzan 1	&	5.21	&	I	& {	85	} &	20	\\
J1735$-$3028G	&	Terzan 1	&	3.92	&	I	& {	85	} &	20	\\
J1737$-$0314B	&	M14 (NGC 6402)	&	8.52	&	B	& {	720	} &	10	\\
J1737$-$0314C	&	M14 (NGC 6402)	&	8.46	&	B	& {	720	} &	10	\\
J1737$-$0314D	&	M14 (NGC 6402)	&	2.89	&	eRB	& {	720	} &	10	\\
J1737$-$0314E	&	M14 (NGC 6402)	&	2.28	&	eRB	& {	720	} &	10	\\
J1748$-$2446C*	&	Terzan 5	&	8.44	&	I	& {	1	} &	21	\\
J1748$-$2446D*	&	Terzan 5	&	4.71	&	I	& {	1	} &	22	\\
J1748$-$2446G*	&	Terzan 5	&	21.67	&	I	& {	3	} &	22	\\
J1748$-$2446I*	&	Terzan 5	&	9.57	&	B	& {	3	} &	22	\\
J1748$-$2446M*	&	Terzan 5	&	3.57	&	B	& {	3	} &	22	\\
J1748$-$2446R*	&	Terzan 5	&	5.03	&	I	& {	3	} &	22	\\
J1748$-$2446S*	&	Terzan 5	&	6.12	&	I	& {	3	} &	22	\\
J1748$-$2446T*	&	Terzan 5	&	7.08	&	I	& {	3	} &	22	\\
J1748$-$2446U*	&	Terzan 5	&	3.29	&	B	& {	3	} &	22	\\
J1748$-$2446W*	&	Terzan 5	&	4.21	&	B	& {	3	} &	22	\\
J1748$-$2446Y*	&	Terzan 5	&	2.05	&	B	& {	3	} &	22	\\
J1748$-$2446aa*	&	Terzan 5	&	5.79	&	I	& {	3	} &	23	\\
J1748$-$2446ab*	&	Terzan 5	&	5.12	&	I	& {	3	} &	23	\\
J1748$-$2446ac*	&	Terzan 5	&	5.09	&	I	& {	3	} &	23	\\
J1748$-$2446ae*	&	Terzan 5	&	3.66	&	B	& {	3	} &	23	\\
J1748$-$2446af*	&	Terzan 5	&	3.3	&	I	& {	3	} &	23	\\
J1748$-$2446ag*	&	Terzan 5	&	4.45	&	I	& {	3	} &	23	\\
J1748$-$2446ah*	&	Terzan 5	&	4.97	&	I	& {	3	} &	23	\\
\end{tabular}
    \label{tab:appendix_table}
\end{table*}

\begin{table*}
    \centering
    \begin{tabular}{lccccc}
\hline
Pulsar Name	&	GC Name	&	Period	&	Type	&	$L_X$(0.3--8 keV)	&	References	\\
 & & (ms) & & {($\times$10$^{30}$ \ergs)} &  \\
\hline
J1748$-$2446ai*	&	Terzan 5	&	21.23	&	B	& {	3	} &	23	\\
J1748$-$2446aj*	&	Terzan 5	&	2.96	&	I	& {	3	} &	24	\\
J1748$-$2446ak*	&	Terzan 5	&	1.89	&	I	& {	3	} &	24	\\
J1748$-$2446al	&	Terzan 5	&	5.95	&	I	& {	3	} &	24	\\
J1748$-$2446am*	&	Terzan 5	&	2.93	&	B	& {	3	} &	25	\\
J1748$-$2446an*	&	Terzan 5	&	4.8	&	B	& {	3	} &	26	\\
J1748$-$2021B*	&	NGC 6440	&	16.76	&	B	& {	160	} &	27	\\
J1748$-$2021C*	&	NGC 6440	&	6.23	&	I	& {	16	} &	27	\\
J1748$-$2021D*	&	NGC 6440	&	13.5	&	eRB	& {	19	} &	27	\\
J1748$-$2021E*	&	NGC 6440	&	16.26	&	I	& {	11	} &	27	\\
J1748$-$2021F*	&	NGC 6440	&	3.79	&	B	& {	17	} &	27	\\
J1748$-$2021G	&	NGC 6440	&	5.21	&	I	& {	260	} &	28	\\
J1748$-$2021H	&	NGC 6440	&	2.85	&	B	& {	260	} &	28	\\
J1801$-$0857A*	&	NGC 6517	&	7.18	&	I	& {	16	} &	29	\\
J1801$-$0857C*	&	NGC 6517	&	3.74	&	I	& {	100	} &	29	\\
J1801$-$0857D*	&	NGC 6517	&	4.23	&	I	& {	30	} &	29	\\
J1801$-$0857E*	&	NGC 6517	&	7.6	&	I	& {	240	} &	30	\\
J1801$-$0857F*	&	NGC 6517	&	24.89	&	I	& {	240	} &	30	\\
J1801$-$0857H*	&	NGC 6517	&	5.64	&	I	& {	30	} &	30	\\
J1801$-$0857I	&	NGC 6517	&	3.25	&	I	& {	240	} &	30	\\
J1803$-$3002A*	&	NGC 6522	&	7.1	&	I	& {	5	} &	31	\\
J1803$-$3002B	&	NGC 6522	&	4.4	&	I	& {	500	} &	32	\\
J1803$-$3002C	&	NGC 6522	&	5.84	&	I	& {	500	} &	32	\\
J1803$-$3002D	&	NGC 6522	&	5.53	&	I	& {	500	} &	33	\\
B1802$-$07*	&	NGC 6539	&	23.1	&	B	& {	16	} &	34	\\
J1807$-$2459B*	&	NGC 6544	&	4.19	&	B	& {	6	} &	36	\\
J1824$-$2452B*	&	M28 (NGC 6626)	&	6.55	&	I	& {	3	} &	37	\\
J1824$-$2452G*	&	M28 (NGC 6626)	&	5.91	&	BW	& {	13	} &	37	\\
J1824$-$2452L	&	M28 (NGC 6626)	&	4.1	&	BW	& {	220	} &	38	\\
J1824$-$2452M	&	M28 (NGC 6626)	&	9.57	&	B	& {	820	} &	39	\\
J1824$-$2452N	&	M28 (NGC 6626)	&	3.35	&	BW	& {	820	} &	39	\\
J1835$-$3259A	&	NGC 6652	&	3.89	&	B	& {	110	} &	40	\\
J1835$-$3259B	&	NGC 6652	&	1.83	&	B	& {	110	} &	41	\\
J1836$-$2354B*	&	M22 (NGC 6656)	&	3.23	&	I	& {	1	} &	42	\\
J1910$-$5959G	&	NGC 6752	&	4.79	&	I	& {	50	} &	43	\\
J1910$-$5959H	&	NGC 6752	&	2.01	&	I	& {	50	} &	43	\\
J1910$-$5959I	&	NGC 6752	&	2.65	&	I	& {	50	} &	43	\\
J1911+0102A*	&	NGC 6760	&	3.62	&	BW	& {	10	} &	44	\\
J1911+0102B*	&	NGC 6760	&	5.38	&	I	& {	10	} &	45	\\
J2133$-$0049A	&	M2 (NGC 7089)	&	10.15	&	B	& {	860	} &	10	\\
J2133$-$0049B	&	M2 (NGC 7089)	&	6.97	&	B	& {	860	} &	10	\\
J2133$-$0049C	&	M2 (NGC 7089)	&	3	&	B	& {	860	} &	10	\\
J2133$-$0049D	&	M2 (NGC 7089)	&	4.22	&	B	& {	860	} &	10	\\
J2133$-$0049E	&	M2 (NGC 7089)	&	3.7	&	B	& {	860	} &	10	\\
J2140$-$2310B	&	M30 (NGC 7099)	&	13	&	B	& {	11	} &	46	\\
\hline
\end{tabular}
    \contcaption{{\it Notes}: *MSPs with precise timing positions. \\
    References (for X-ray limits, and pulsar properties\footnote{See \url{http://www.naic.edu/~pfreire/GCpsr.html} for up-to-date radio pulsar properties}): 
    1) \citet{Manchester91,Robinson95,Freire01,Heinke2005,Freire17};
    2) \citet{Camilo00,Ridolfi2016,Heinke2005};
    3) \citet{Heinke2005,Ridolfi2021};
    4) \citet{Pan2021b} and this work;
    5) \citet{Henleywillis2018,Dai2020};
    6) \citet{Hessels07,Zhao2019,Pan2021b,Qian21}, this work;
    7) \citet{Anderson1997}, this work; 
    8) \citet{Hessels07,Pallanca2014}, this work;
    9) \citet{Hessels07,Bahramian2020}; 
    10) \citet{Bahramian2020,Pan2021b};
    11) \citet{Bahramian2020}, Pan et al. in prep\footnote{https://crafts.bao.ac.cn/pulsar/SP2/};
    12) \citet{Kulkarni1991,Wang2020,Zhao2021}; 
    13) \citet{Lu2009}, Pan \& FAST team in prep \footnote{https://fast.bao.ac.cn/cms/article/65/};
    14) \citet{Pan2021b,Bahramian2020};
    15) \citet{DAmico01,Possenti2003,Lynch2012}, this work;
    16) \citet{Lynch2012}, this work;
    17) \citet{Ridolfi2021,Oh2020};
    18) \citet{Lu2011,Pan2020,Pan2021b}, this work; 
    19) TRAPUM in prep \footnote{http://trapum.org/discoveries.html}, this work;
    20) \citet{Cackett2006}, DeCesar et al. in prep;
    21) \citet{Lyne00,Prager17,Bogdanov2021};
    22) \citet{Ransom05,Prager17,Bogdanov2021};
    23) \citet{Prager17,Bogdanov2021};
    24) \citet{Prager17,Cadelano18,Bogdanov2021},
    25) \citet{Andersen18,Bogdanov2021};
    26) \citet{Bogdanov2021,Ridolfi2021};
    27) \citet{Freire2008}, this work;
    28) \citet{Pooley2002}, TRAPUM in prep, this work;
    29) \citet{Lynch2011}, this work;
    30) \citet{Pan2021a}, this work;
    31) \citet{Possenti05,Bahramian2020,Zhang2020}, this work; 
    32) \citet{Bahramian2020}, GBT team in prep;
    33) \citet{Bahramian2020,Ridolfi2021};
    34) \citet{D'Amico1993,Thorsett1993}, this work;
    35) \citet{DAmico01,Ransom2001,Lynch2012}, this work; 
    36) \citet{Lynch2012,Bahramian2020}, this work;
    37) \citet{Begin06,Bogdanov2011};
    38) \citet{Bogdanov2011}, GBT team in prep;
    39) \citet{Bahramian2020}, TRAPUM in prep;
    40) \citet{Stacey2012,DeCesar2015};
    41) \citet{Stacey2012}, Gautam et al. in prep;
    42) \citet{Lynch2011,Bahramian2020}, this work;
    43) \citet{Forestell2014}, TRAPUM in prep;
    44) \citet{Deich1993,Freire2005,Bahramian2020}, this work; 
    45) \citet{Freire2005,Bahramian2020}, this work;
    46) \citet{Ransom04,Zhao2020b}; 
    }

\end{table*}




\bsp	
\label{lastpage}
\end{document}